\documentclass[]{pasj01}
\usepackage{bm}
\usepackage{subfigure}
\usepackage{lineno}
\usepackage{multicol}
\usepackage{graphicx}
\usepackage{epstopdf}
\usepackage{ulem}

\Received{2019/12/14}
\Accepted{2020/03/17}

\begin{document} 
\title{Variations of the physical parameters of the blazar Mrk~421
 based on the analysis of the spectral energy distributions}

\author{Yurika \textsc{Yamada}\altaffilmark{1}}
\altaffiltext{1}{Department of Physical Science, Hiroshima University,
1-3-1 Kagamiyama, Higashi-Hiroshima, Hiroshima 739-8526, Japan}

\author{Makoto \textsc{Uemura}\altaffilmark{2}}
\altaffiltext{2}{Hiroshima Astrophysical Science Center, Hiroshima
  University, 1-3-1 Kagamiama, Higashi-Hiroshima, 739-8526, Japan}
\email{uemuram@hiroshima-u.ac.jp}

\author{Ryosuke \textsc{Itoh}\altaffilmark{3}}
\altaffiltext{3}{Department of Physics, Tokyo Institute of Technology,
  2-12-1 Ohokayama, Meguro, Tokyo 152-8551, Japan}

\author{Yasushi \textsc{Fukazawa}\altaffilmark{1}}

\author{Masanori \textsc{Ohno}\altaffilmark{1}}

\author{Fumiya \textsc{Imazato}\altaffilmark{1}}


\maketitle

\begin{abstract}

We report on the variations of the physical parameters of the jet
observed in the blazar Mrk~421, and discuss the origin of X-ray flares
in the jet, based on the analysis of the several spectral energy
distributions (SEDs). The SEDs are modeled using the one-zone
synchrotron self-Compton (SSC) model and its parameters determined
using a Markov chain Monte Carlo method. The lack of data at TeV
energies means many of the parameters cannot be uniquely determined
and are correlated. These are studied in detail. We found that the
optimal solution can be uniquely determined only when we apply a
constraint to one of four parameters: the magnetic field ($B$),
Doppler factor, size of the emitting region, and
normalization factor of the electron energy distribution. We used 31
sets of SED from 2009 to 2014 with optical--UV data observed with
UVOT/$\it{Swift}$ and the Kanata telescope, X-ray data with
XRT/$\it{Swift}$, and $\gamma$-ray data with the $Fermi$ Large Area
Telescope (LAT). The result of our SED analysis suggests that, in the
X-ray faint state, the emission occurs in a relatively small area
($\sim 10^{16}\;{\rm cm}$) with relatively strong magnetic field
($B\sim 10^{-1}\;{\rm G}$). The X-ray bright state shows a tendency
opposite to that of the faint state, that is, a large emitting area
($\sim 10^{18}\;{\rm cm}$), probably in the downstream of the jet and
weak magnetic field ($B\sim 10^{-3}\;{\rm G}$). The high X-ray flux
was due to an increase in the maximum energy of electrons. On
the other hand, the presence of two kinds of emitting areas
implies that the one-zone model is unsuitable to reproduce, at least a 
part of the observed SEDs.
\end{abstract}

\section{Introduction}

Blazars are a type of active galactic nuclei (AGN) whose jets are
directed toward us. The radiation of the jet dominates the entire
spectral energy distribution (SED) due to a strong beaming effect
(\cite{Urry}). This effect also causes rapid variability and high
luminosity, commonly observed in blazars. Hence, blazars are good
candidates for studying the physical processes of relativistic jets.

The SED of blazars extends from the radio to the $\gamma$-ray bands
(\cite{Fossati}). They exhibit two broad components: a low energy
component, attributed to synchrotron radiation from relativistic
electrons, and a high energy component, generally believed to be
inverse Compton (IC) scattering by electrons. The local synchrotron
radiation can act as seed photons for the IC process. This scenario is
called the synchrotron self-Compton (SSC) process
(\cite{Tavecchio1998}). The modeling of SED with a given radiation
mechanism, like the SSC process, allows us to investigate the
intrinsic physical properties of the emitting regions of the blazar
and the physical conditions of the jet (e.g., \cite{Ghisellini2009};
\cite{Bottcher2007}).

The SSC model can be expressed with about $10$ physical parameters,
including the magnetic field, Doppler factor, and parameters of the
electron energy distribution (e.g., \cite{Finke2008}). It is difficult
to determine the unique solution of the problem because some of the
parameters are strongly correlated and the model is degenerate. In
previous studies, the parameters were estimated by fixing the size of
the emitting region (e.g., \cite{Bartoli2016}), Doppler factor,
magnetic field, or all of them (e.g., \cite{Tramacere2009};
\cite{itoh2015}). The model optimization procedure is sometimes based
on a ``fit-by-eye'' method, with which a real solution can be easily
overlooked, in particular if  the model is degenerate
(e.g., \cite{Bottcher2013}).

The optimal solution may be misidentified if the parameters are set to
unsuitable values. In evaluating the likelihood function or posterior
probability density function in a multidimensional space, it is
essential to understand the degenerate structure of the model and
thereby to establish a method to estimate the optimal solution. The
likelihood function also allows the uncertainty of the parameters to
be estimated, which is especially important for considering the
variations of the model parameters.

Mrk~421 ($z = 0.031$) is a well-known nearby blazar (\cite{Punch1992}).
It is a very active source, exhibiting major outbursts about once
every two years composed of many short flares of both X-rays and
$\gamma$-rays (\cite{Aielli2010}; \cite{Bartoli2011a}). The SED of
Mrk~421 can be modeled with a simple one-zone SSC model
(\cite{Paggi2009}; \cite{Abdo2011}). Thus, Mrk~421 is an ideal object
to study the variations of the SSC model parameters, and thereby to
understand the physical process in the jets. 

In this paper, we use the Markov chain Monte Carlo (MCMC) method to 
study the degenerate structure of the SSC model and a proper
methodology for fitting the model to the SED data. \citet{gia15MMML}
propose a method to estimate the model parameters not from the
prepared SED data, but from the count-based data obtained with each
instrument. Since the SED data is usually prepared through a fitting
procedure to each count-based data, the method provides a better way
to achieve the true maximum likelihood solution. The present paper
focuses on the classical analysis with the SED data, whereas the
degenerate structure of the model is common in both methods. 
We investigate the variation of the physical parameters of
the jet using the SED data of Mrk 421 from 2009 to 2014, which
includes a bright X-ray outburst in 2010. The paper is organized as
follows. In Section 2 we introduce the dataset. In Section 3 we
describe the model and MCMC method. We demonstrate that we need to
restrict one parameter to uniquely determine the optimal solution. In
this paper, we set two kinds of restrictions: on the size of the
emitting region (or the variation time-scale), and
on the Doppler factor. In Section 4 we report on a time-series
analysis for the estimation of the variation time-scale of the
object. The results of the SED analysis are presented in Section 5,
followed by a discussion of the origin of the X-ray variation in
Section 6. Section 7 is a summary.

\section{Observations}
\subsection{X-ray data}

The {\it Neil Gehrels Swift} observatory is a NASA mission, launched
in 2004, devoted to observing fast transients, namely prompt and
afterglow emission of $\gamma$-ray bursts \citep{Gehrels2004}.
The satellite carries three sets of instruments: the Burst Alert
Telescope (BAT; 15--150 keV; \cite{Gehrels2004}), the X-Ray Telescope
(XRT; 0.3--10 keV; \cite{Burrows2005}), and the Ultra-Violet Optical
Telescope (UVOT; 170--650 nm; \cite{Roming2005}). In this paper, we
analyzed the data of Mrk 421 taken by $\it{Swift}$ between 2009 and
2014. 

\begin{figure}
\begin{center}
\subfigure{\includegraphics[width=5.5cm,angle=-90]{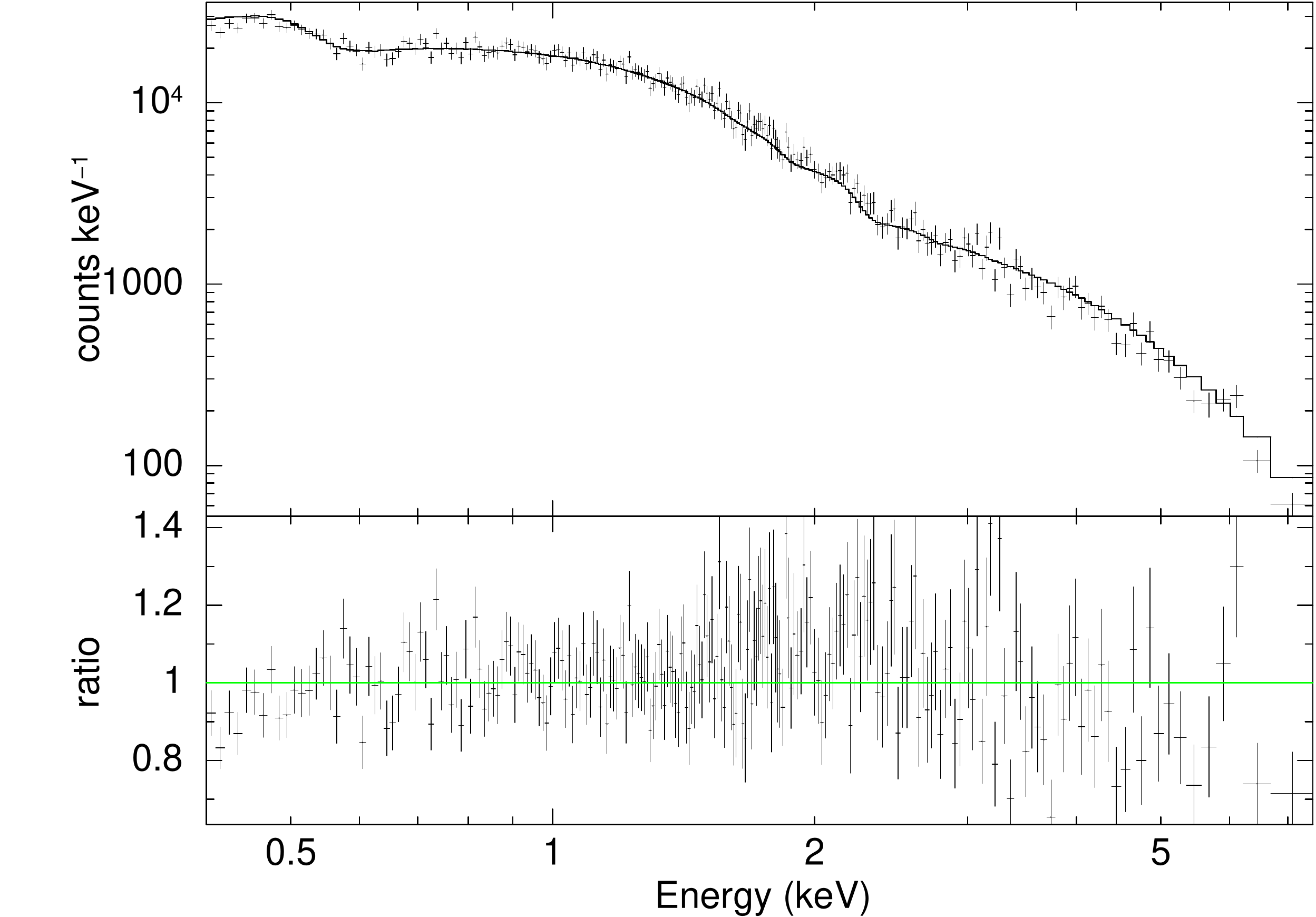}
 \label{fig:fig01left}}
\subfigure{\includegraphics[width=5.5cm,angle=-90]{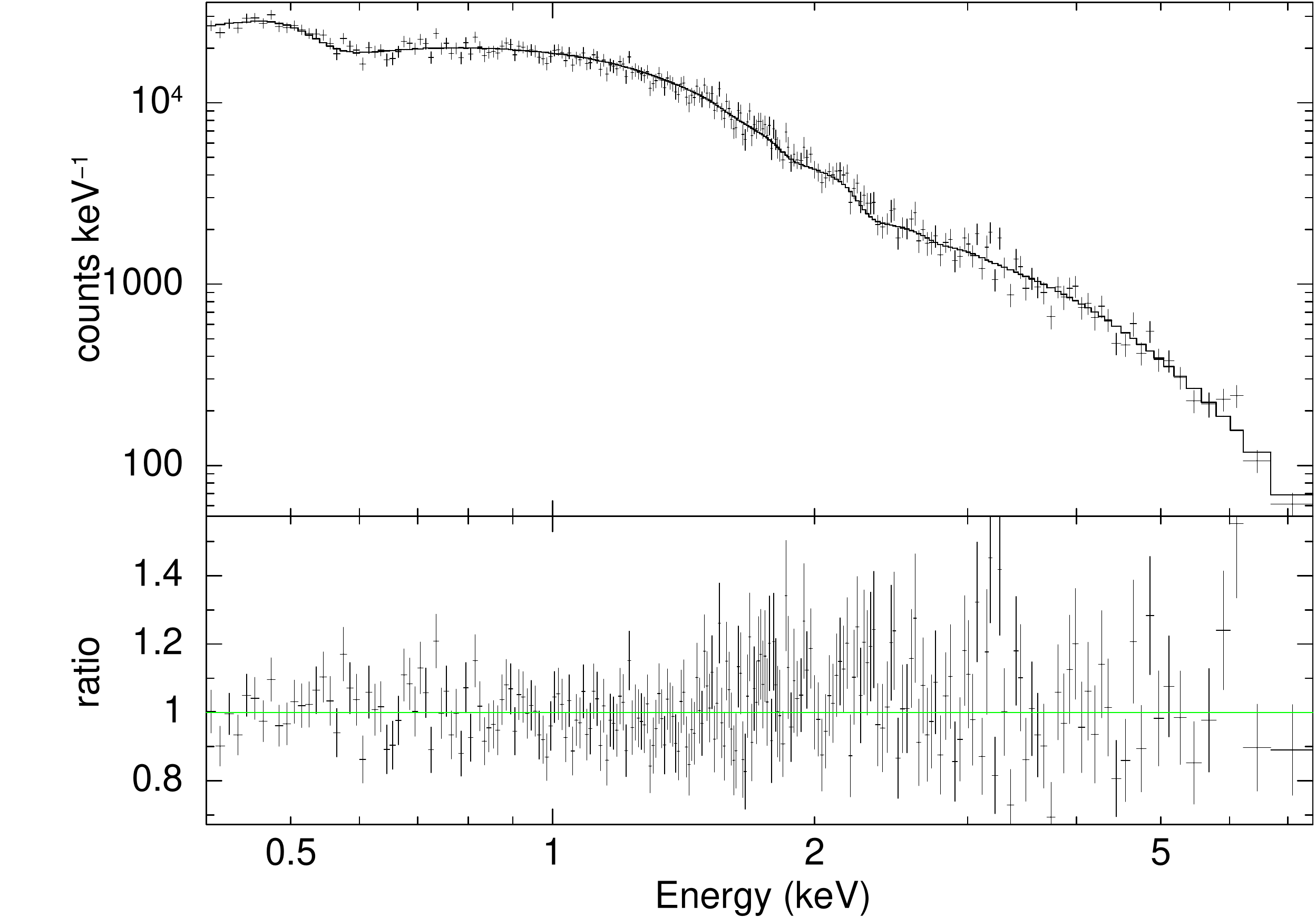}
 \label{fig:fig01right}}
\end{center}
\caption{Examples of the X-ray spectra and models of Mrk~421. The data
  shown in the figure were taken by {\it Swift}/XRT on MJD~55186. For
  each panel, the ratio of the data to model is also shown. The
  power-law and log-parabolic models are used in the upper and lower
  panels, respectively.}\label{fig:spe}
\end{figure}

The $\it{Swift}$-XRT is a Wolter type-I grazing incidence telescope
that focuses X-rays onto a CCD. This instrument has a 110 $\rm{cm}^2$
effective area, 23.6 arcmin FOV, and 15 arcsec angular resolution.
The XRT data of Mrk 421 used in this paper were all taken in windowed
timing mode, due to the high flux rate of the source. The exposure
time of each run is typically $10^3\;{\rm s}$. The XRT data
were reduced using the software distributed with the HEASoft v6.19
package by NASA High Energy Astrophysics Archive Research Center
(HEASARC). Spectral analysis was performed with XSPEC version
12.9.0o. We fitted the XRT spectra with a log-parabolic or power-law
shape component and the Galactic absorption component represented by
the {\tt wabs} model over an energy range of 0.3--10~keV. The column
density of the absorbing component was fixed to
$1.61\times 10^{20}\,{\rm cm}^{-2}$ \citep{Lockman1995}. We confirmed 
that the $\chi^2$ values for the power-law model were larger than
those for the log-parabolic model for all data sets. The difference in
$\chi^2$ was at least 3, and mostly $>50$. Examples of the X-ray
spectra are shown in Figure \ref{fig:spe}. The residuals between the
data and power-law model are systematically large both in $\lesssim
0.5$~keV and $\gtrsim 5$~keV, compared with those of the log-parabolic
model. Hence, all data was analyzed with the log-parabolic model. In
the SED fitting performed in Section 3 and 5, we use the absorption
corrected XRT data. 

\begin{figure*}
 \begin{center}
   \includegraphics[width=17.0cm]{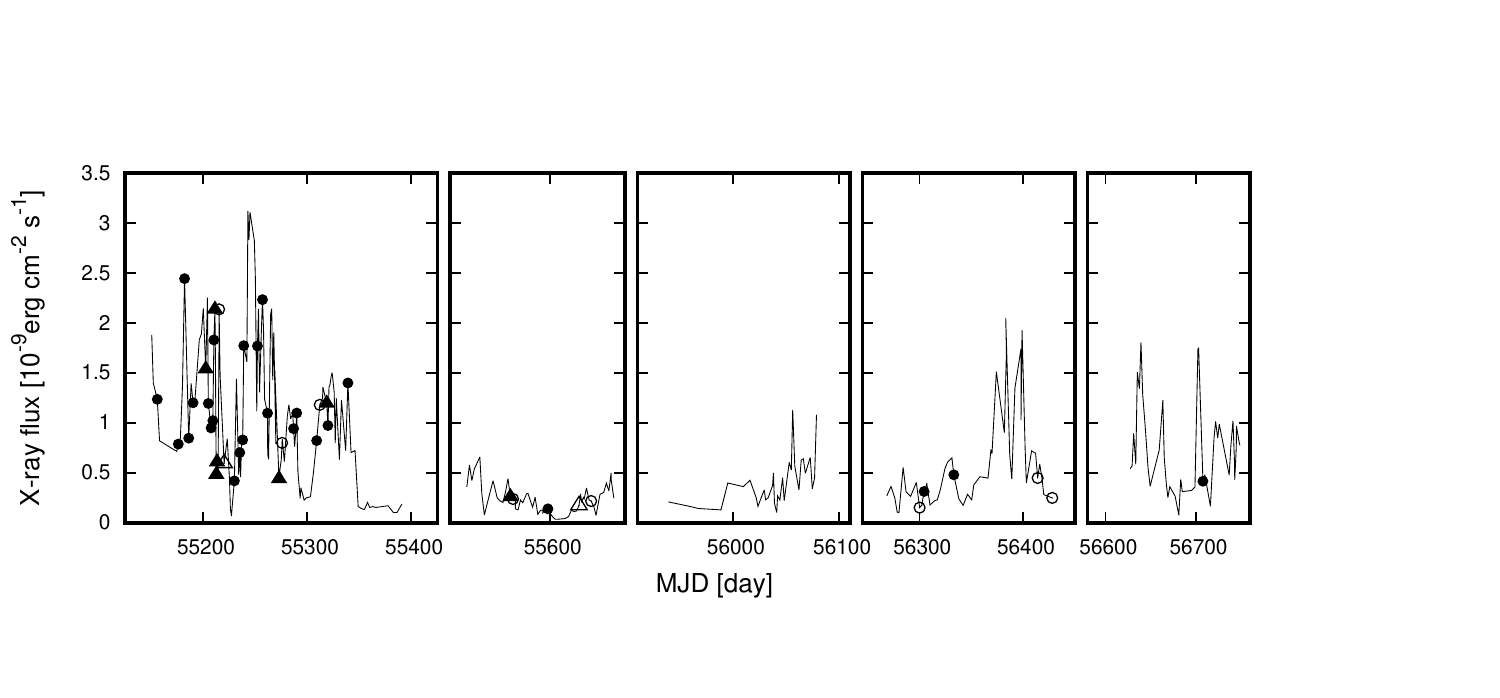}
\end{center}
    \caption{X-ray light curve of Mrk 421 for 2009--2014 obtained with
    XRT/$\it{Swift}$. The gray line indicates the light curve of all
    data. The filled circles and triangles indicate the epochs at
    which we report the estimated SED parameters in this paper. The
    triangles show the epochs at which the optical and near-infrared
    Kanata observations are available. The open symbols indicate the
    epochs of the outlier SEDs (see section~3.3). The measurement
    error is smaller than the symbol size.}\label{fig:lightcurve} 
\end{figure*}

Figure \ref{fig:lightcurve} shows the X-ray light curve of Mrk 421
observed with XRT from 2009 to 2014. The data were acquired with a
sampling interval of about a day over 100 days in each year.
The object experienced an outburst in 2009--2010 (MJD 55150--MJD
55340). After the outburst ended, it was faint for most of the
observation period (X-ray flux
$< 0.5\times 10^{-9}\; \rm{erg\, cm^{-2}\, s^{-1}}$ at 0.310 keV),
while it occasionally experienced short flares.
In both the bright and faint states, the object significantly
fluctuates in X-ray flux over a day, which would suggest that $< 1$
day variations are likely. The features of the X-ray light curves of
Mrk 421 have already been presented in past studies (e.g.,
\cite{Bartoli2016}).

Due to the long calculation time ($\sim 10$ hours) to obtain $10^5$
MCMC samples for each SED dataset, we refrained from analyzing all the
available SEDs. In this study, we focus on which SSC parameters change
and how they change depending on the X-ray flux in order to study
the origin of the X-ray variations. Hence, we selected the data to
cover various X-ray levels. First, we selected the SED data which
include the optical data observed by the Kanata telescope. Second, we
randomly selected dates in the X-ray bright state (X-ray flux
$> 0.5 \times 10^{-9}\; \rm{erg\, cm^{-2}\, s^{-1}}$) during the 2010
outburst at time intervals of $> 2$ days. In the faint state, we
randomly selected data between 2009 and 2014. We chose 28 days for the
bright state and 13 days for the faint state for our SED analysis.
The epochs of the XRT data used in the paper for the SED analysis are
listed in Appendix 1.

\subsection{$\gamma$-ray data}

The $Fermi$ Large Area Telescope (LAT) is a $\gamma$-ray telescope
which covers the energy range from 20 MeV to more than 300 GeV
\citep{Atwood2009}. We analyzed the LAT Pass 8 data (P8R2) for Mrk 421
\citep{Atwood2013}. Source class events (event type 3, event class
128: front and back events) were selected for zenith angles $<$
$90^{\circ}$ with the tool {\tt gtselect}, 
and a time region was selected with the filter
expression {\tt (DARA$\_$QUAL$>$0) $\& \&$ (LAT$\_$CONFIG==1)}.
The analysis was performed using the Science Tools version v10r0p5
provided by the $\it{Fermi}$-LAT collaboration. We used the gtlike
tool, based on a binned maximum likelihood method to build the SED of
the source. The binned analysis is performed with a spatial
bin size of a square of $0.1 \times 0.1\;{\rm deg}^2$ in the area of
$20 \times 20\;{\rm deg}^2$ square map region centered on Mrk 421. The
field background point sources within 25 deg from Mrk~421,
listed in the LAT Third Source Catalog (3FGL; \cite{Acero2015}), were
all included, and their spectra were assumed to be the same model as
in the catalog. The model parameters related to the spectral shape 
were fixed to the catalog values for all sources.
The model normalization was set to be free for sources
$< 10^\circ$ and fixed for sources $>10^\circ$ of Mrk~421. The
spectral analysis of the resulting data set was carried out by
including the galactic diffuse emission component model
({\tt gll$\_$iem$\_$v06.fits}; \cite{Acero2016}) and an isotropic
background component model
({\tt iso$\_$P8R2$\_$SOURSE$\_$V6$\_$v06.txt}) with a post-launch 
instrumental response function {\tt P8$\_$R2$\_$SOURCE$\_$V6}.
The normalizations of these two models were left free in the following 
analysis. 

We derived the GeV $\gamma$-ray spectrum by performing a likelihood
analysis for six energy bands, logarithmically spaced from 100 MeV to
300 GeV. The spectral studies with the likelihood analysis were
performed for the 41 epochs of our sample defined in section 2.1. We
extracted five days before and after the observation data of
$\it{Swift}$. We used  the power-law model to describe the source
spectrum described by the expression $f(E) = KE^{\alpha}$, where $K$
and $\alpha$ indicate the normalization flux and the spectral index.

The integration time of the $\gamma$-ray data (10 days) is
much longer than those of X-ray and optical data used in this
paper. Using the 10-d averaged $\gamma$-ray flux is unavoidable in
order to obtain meaningful data of this $\gamma$-ray faint
source. Using the 10-d averages mean that we assume the $\gamma$-ray
flux to not vary significantly in a time-scale of days, but
possibly in a time-scale of a few tens of days. According to
\citet{Bartoli2016} which present a detailed report of the long-term
$\gamma$-ray light curve of Mrk 421, the assumption is almost
consistent with the observed variation. \citet{Bartoli2016} reported a
few short flares having a duration of a few days. The 10-d average
flux would definitely smear them out. As a result, we may
underestimate the true $\gamma$-ray flux at the epochs of the short
$\gamma$-ray flares. However, the flare amplitude is not very large,
by a factor of $\sim 3$ at maximum. The $\gamma$-ray flux level
affects, in particular, the estimation of $T$. In this paper, we only
discuss variations in $T$ over an order of magnitude, and do not
discuss its minor variation. 

\subsection{Optical and ultraviolet data}

We used the optical data of Mrk 421 from 2009 to 2010
reported by \citet{itoh2015}. The data is the $V$ and $R_{\rm c}$ band
photometric data obtained with the HOWPol instrument installed on the
1.5-m Kanata telescope located at the Higashi-Hiroshima Observatory,
Japan \citep{Kawabata2008}. 

We also obtained UV data with $\it{Swift}$-UVOT. For UVOT
observations, we used three ultraviolet filters in the imaging mode,
with effective central wavelengths of 260.0 nm (UVW1), 224.6 nm
(UVM2), and 192.8 nm (UVW2). The UVOT data were reduced following the
standard procedure for CCD photometry. The counts were extracted from
an aperture of 5'' radius for all filters and converted to flux using
the standard zero points \citep{Poole2008}. For these reductions, we
used UVOTSOURCE from HEADAS 6.19 and the calibration database CALDB
released on 17 July 2015. Galactic extinction correction was performed
using the method of \citet{Cardelli1989} with $E(B-V) = 0.0132$ from
\citet{Schlafly2011}. 

\section{Method}
\subsection{Synchrotron self-Compton (SSC) model}

In this study, we used the one-zone SSC model for the observed SED
(\cite{Tavecchio1998}; \cite{Fossati2008}). This model assumes that
high energy electrons emitting synchrotron photons upscatter the
photons by inverse Compton scattering. We calculated the SED based on
the SSC formula given by \citet{Finke2008}. This model calculates the
synchrotron and inverse Compton radiation from a region in which
electrons are confined and have an isotropic energy distribution,
$N_{\rm e}(\gamma)$, where $\gamma$ is the Lorentz factor of the
electrons. The model SED is a function of the magnetic field $B$,
Doppler factor $\delta_{\rm{D}}$, light crossing time of the
emitting region $T$, and $N_{\rm{e}}(\gamma)$. A comoving region
size, $R$ is calculated from $T$ and $\delta_{\rm{D}}$ as 
$R=c \delta_{\rm{D}} T /(1+z)$. 
We used a broken power-law form for $N_{\rm{e}}(\gamma)$:
\begin{eqnarray}\label{eq:1}
N_{\rm{e}} (\gamma)=K_{\rm{e}} \times\left\{ \begin{array}{ll}         
\left(\frac{\gamma}{\gamma_{\rm{b}}}\right)^{-p_0} & (\gamma_{\rm min} < \gamma < \gamma_{\rm b})  \\
\left(\frac{\gamma}{\gamma_{\rm{b}}}\right)^{-p_1} & (\gamma_{\rm b} < \gamma < \gamma_{\rm max}) \\
\end{array} \right.                                                                            \end{eqnarray}
where $K_{\rm{e}}$ is the electron normalization factor,
$\gamma_{\rm{b}}$ is the break energy, $p_0$ and $p_1$ are the
electron spectral indices, and $\gamma_{\rm min}$ and
$\gamma_{\rm max}$ are the minimum and maximum $\gamma$. We
used exact synchrotron and full Klein-Nishina expressions, while
\citet{Finke2008} propose approximation methods for them. The
$\gamma\gamma$ photoabsorption was not included in the model.

\citet{Abdo2011} found that, in order to properly describe the shape
of the measured broadband SED of Mrk 421, the model requires an
electron distribution parameterized with three power law functions
(and hence two breaks). A high energy break is required to reproduce
the observed hard X-ray data in the range 20--150 keV. Our study did
not include the hard X-ray data in the SED analysis. Therefore, the
broken power law model of equation (\ref{eq:1}) is sufficient for our
data sets. 

We found several cases in which $p_1$ diverged to a large
value and terminate the MCMC process due to overflow. In such cases,
the data required a lower limit of $p_1$ ($p_1>5$), while an upper
limit was not given. In this paper, we fixed $p_1$ to be a large
value, $p_1=20$ in those cases. Furthermore, we confirmed that the
SED data in this study cannot constrain $\gamma_{\rm{min}}$ and
$\gamma_{\rm{max}}$. $\gamma_{\rm {b}}$ is estimated to be $10^4-10^7$
as shown in Section 5. Hence, $\gamma_{\rm min}$ and
$\gamma_{\rm max}$ do not affect the estimation if
$\gamma_{\rm min}<10^4$ and $\gamma_{\rm max}>10^7$. We fix
$\gamma_{\rm{min}}$ and $\gamma_{\rm{max}}$ as $10^2$ and $10^8$ in
our analysis.

\subsection{MCMC algorithm}

MCMC is a method for simulating random samples from probability
distributions. The posterior probability distribution gives the
optimal values of parameters and their uncertainties. We estimate the
posterior probability distributions of the SSC model parameters using
MCMC. Here, we consider the model parameters, $\bm{x}$, and the data,
$\bm{y}$. According to the Bayes' theorem, the posterior distribution
of the parameters $P(\bm{x}|\bm{y})$ is proportional to the
likelihood function $L(\bm{y}|\bm{x})$ and prior distribution
$P(\bm{x})$, as follows:
\begin{equation}\label{eq:2}
P(\bm{x}|\bm{y})\propto L(\bm{y}|\bm{x})P(\bm{x}) 
\end{equation} 
We define the likelihood function by the following normal distribution:
\begin{equation}\label{eq:5}
L= \left \{
\begin{array}{l}
\prod_{i}\frac{1}{\sqrt{2\pi\sigma^2_{d,i}}}\exp\left(-\frac{(d_i(\nu)-f^{\mathrm{syn}}_{i})^2}{2\sigma^2_{d,i}}\right) (\nu<10^{21}\,\mathrm{Hz})\\ 
\prod_{i}\frac{1}{\sqrt{2\pi\sigma^2_{d,i}}}\exp\left(-\frac{(d_i(\nu)-f^{\mathrm{SSC}}_{i})^2}{2\sigma^2_{d,i}}\right) (\nu>10^{21}\,\mathrm{Hz})
\end{array}
\right.
\end{equation} 
where $f^{\mathrm{syn}}$ and $f^{\mathrm{SSC}}$ are the model SED values
of the synchrotron and inverse Compton components, respectively, and
$d_i$ and $\sigma_d$ are the observed SED data and their measurement
uncertainties.

For the SED modeling of blazars, the prior distributions of parameters
may be given by earlier estimations from past studies. For example,
$T$ can be the minimum time-scale of variations, and estimated
from the light curve with its uncertainty. In this case, we can
assume a prior distribution of $T$ from those estimations.

We used the adaptive Metropolis algorithm to sample the posterior
distribution \citep{Haario2001}. This method determines the
variance--covariance matrix of the proposed distribution from MCMC
samples, and enables efficient sampling. The variance--covariance
matrix is updated from step $n-1$ to $n$, as follows:
\begin{eqnarray}\label{eq:4}
\bm{\mu}_n\leftarrow\bm{\mu}_{n-1}+h_n(\bm{x}_n-\bm{\mu}_{n-1}) \nonumber \\
\Sigma_{n}\leftarrow\Sigma_{n-1}+u_n\left((\bm{x}_n-\bm{\mu}_{n-1})(\bm{x}_n-\bm{\mu}_{n-1})^T-\Sigma_{n-1}\right)\\
\sigma^2{_n}\leftarrow\sigma^2_{n-1}+s_n(FA_n-\alpha) \nonumber
\end{eqnarray}
where $\bm{\mu}$ is the mean value vector of $\bm{x}$, $\Sigma$ is the
variance--covariance matrix, and $\sigma^2$ is the scale
parameter. $FA$ is 1 when the candidate of the next state is accepted
and 0 when it is not. $\alpha$ is the acceptance rate. In this study,
we set $\alpha = 0.234$, an optimal acceptance rate of the
Metropolis algorithm (\cite{Roberts1997}). $h_n$, $u_n$, and $s_n$
are learning coefficients in the Robbins-Monro algorithm; we
set a standard form, $h_n = u_n = s_n = \frac{10}{n+100.0}$
(\cite{Robbins1951}). We found that $10^5$ iterations are sufficient
for convergence, discarding the first $2\times10^4$ steps as
burn-in. We estimated the posterior distributions of $B$,
$\delta_{\rm{D}}$, $T$, $K_{\rm{e}}$, and $\gamma_{\rm{b}}$ on a
logarithmic scale ($\log_{10}$, hereafter simply referred as
$\log$) for an efficient sampling in a wide range of parameters, and
those of $p_0$ and $p_1$ on a linear scale.

\subsection{The degenerate structure of the model}

\begin{figure}
 \begin{center}
    \includegraphics[width=8.0cm]{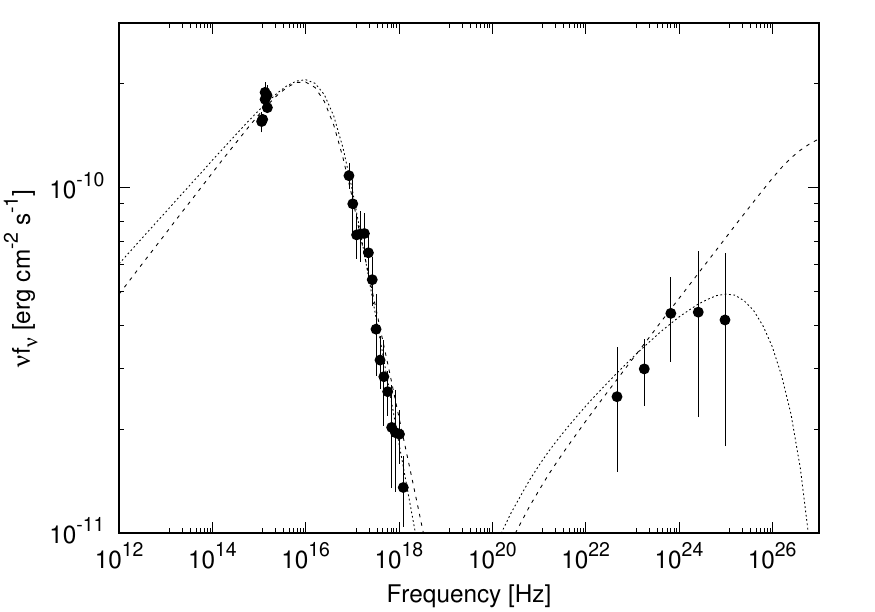}
\end{center}
    \caption{SED of Mrk 421 on MJD 55598 with the SSC model optimized
      by MCMC. The dotted and dashed lines indicate the model
      SEDs having a similar posterior probability. The model
      parameters of the dotted and dashed lines are as follows:
      ($\log B$, $\log\delta_{\rm{D}}$, $\log T$, $\log K_{\rm{e}}$,
      $\log \gamma_{\rm{b}}$, $p_0$, $p_1$) = ($-38$, $38$, $-70$,
      $-31$, $3.8$, $2.2$, $4.4$) and ($-2$, $2.5$, $2$, $41$, $5.1$,
      $2.6$, $4.5$), respectively.}\label{fig:sedAB}   
\end{figure}

First, we set non-informative priors for all parameters, in order to
investigate the degenerate structure of the model. We used uniform
distributions with minimum and maximum values as priors to avoid
overflow of the calculations. The minima and maxima of each parameter
were set to be $\log B=(-40,5)$, $\log \delta_{\rm{D}}=(-5,45)$,
$\log T=(-90,20)$, and $\log K_{\rm{e}}=(-50,50)$. In this section, we
show the data and results for the SED data of Mrk 421 on MJD 55598, as
an example. The observed SED is shown in Figure \ref{fig:sedAB}. 

\begin{figure*}
  \begin{center}
    \includegraphics[bb=0 0 1152 1152,width=16cm]{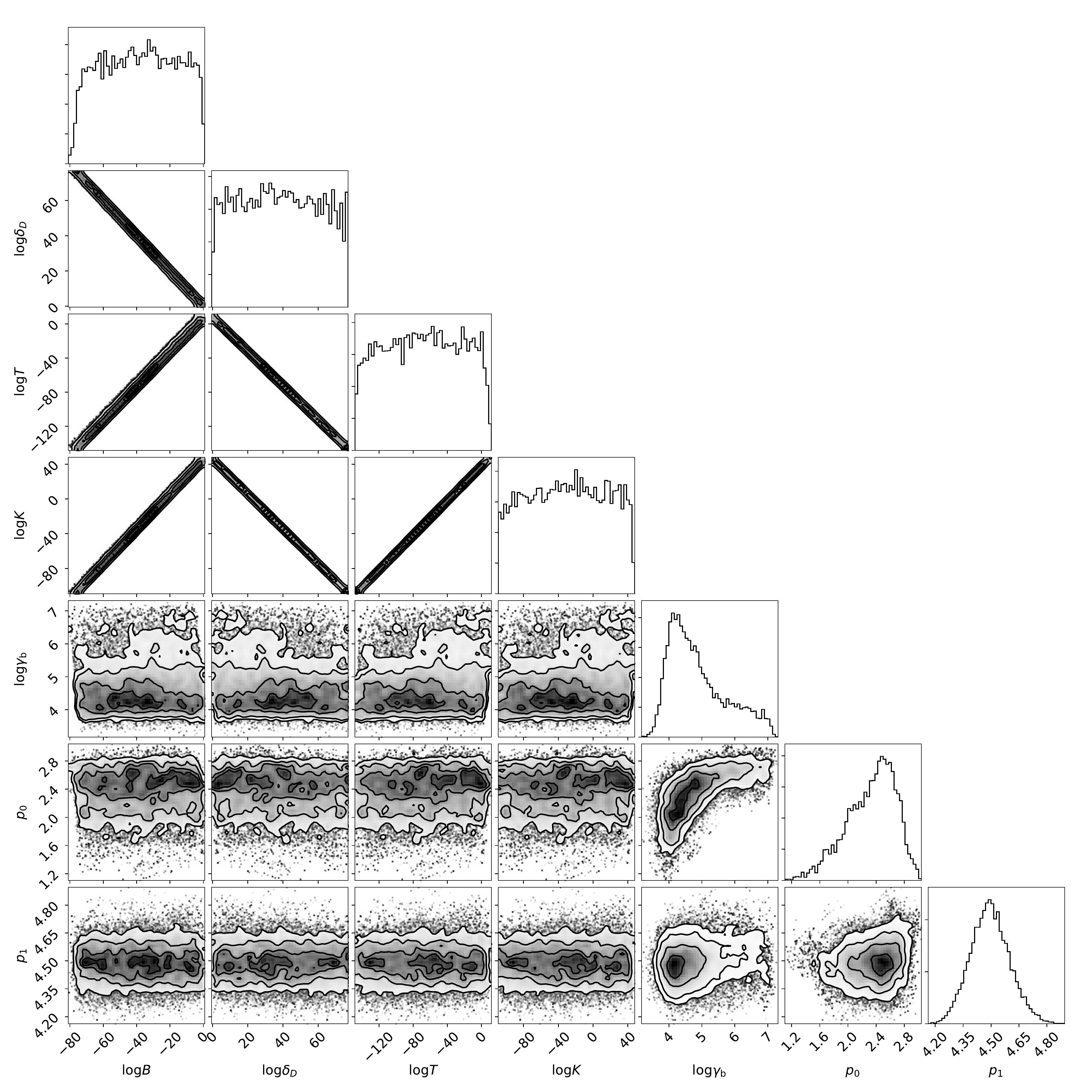}
\end{center}
    \caption{The one- and two-dimensional projections of the posterior
      probability distributions of seven parameters, $\log B$,
      $\log \delta_{\rm{D}}$, $\log T$, $\log K_{\rm{e}}$,
      $\log \gamma_{\rm{b}}$, $p_0$, $p_1$, obtained from the MCMC
      sample shown in Figure \ref{fig:notjizen_trace}.}\label{fig:notjizenhist} 
\end{figure*}

Figure \ref{fig:notjizenhist} shows their posterior probability
distribution obtained from the MCMC sample. The MCMC samples
converge to a stationary distribution as described in Appendix~2.
As can be seen in Figure \ref{fig:notjizenhist}, $\log B$,
$\log \delta_{\rm{D}}$, $\log T$, and $\log K_{\rm{e}}$ have a uniform
probability distribution over a wide range. Table
\ref{tab:soukan} shows the Pearson's correlation coefficients between
the parameters. The coefficients were calculated from obtained MCMC
samples. We can see strong correlations between $\log B$,
$\log \delta_{\rm{D}}$, $\log T$,  and $\log K_{\rm{e}}$ in this table.
In this paper, we define a significant correlation with the statistical 
significance at $> 99\%$ confidence level. As can be seen from the
2D posterior distributions in Figure \ref{fig:notjizenhist}, we can
confirm that those four parameters have a linear correlation in the 4D
parameter space over a wide scale. Therefore, we need to apply a
constraint on any one of $\log B$, $\log\delta_{\rm{D}}$, $\log T$,
and $\log K_{\rm{e}}$ in order to uniquely determine the optimal solution.

The dotted and dashed lines of Figure \ref{fig:sedAB} show the model 
SEDs with the highest likelihood. The parameters for the dotted and
dashed models are very different: ($\log B$, $\log \delta_{\rm{D}}$,
$\log T$, $\log K_{\rm{e}}$, $\log \gamma_{\rm{b}}$, $p_0$, $p_1$) $=$
($-38$, $38$, $-70$, $-31$, $3.8$, $2.2$, $4.4$) and ($-2$, $2.5$,
$2$, $41$, $5.1$, $2.6$, $4.5$), respectively. Both of them reproduce
the observed data points in the optical, X-ray, and GeV $\gamma$-ray
regimes, whereas their difference is the largest in the TeV $\gamma$-ray
regime ($\sim 10^{26}\;{\rm Hz}$). This means that the optimal
solution can be uniquely determined if the TeV data is available. In this 
study, we focus on the variation of the parameters, and hence
simultaneity of the multi-frequency data is necessary. However,
simultaneous TeV data are not available for the dates considered here.
Thus, we need an informative prior of the parameters to obtain a
unique solution. 

Table \ref{tab:soukan} shows that $\log \gamma_{\rm{b}}$, $p_0$, and
$p_1$ have only small correlation coefficients with $\log B$,
$\log \delta_{\rm{D}}$, $\log T$, and $\log K_{\rm{e}}$. As can be
seen in Figure \ref{fig:notjizenhist}, the posterior probability 
distributions of $\log \gamma_{\rm b}$, $p_0$, and $p_1$ are not
uniform, but have single-peaked profiles. Therefore, the optimal
solution can be obtained for the parameters that determine the shape
of the electron energy distribution, such as $\gamma_{\rm b}$, 
$p_0$, and $p_1$, without any informative priors.

\begin{figure}
  \begin{center}
   \includegraphics[clip,width=8.0cm]{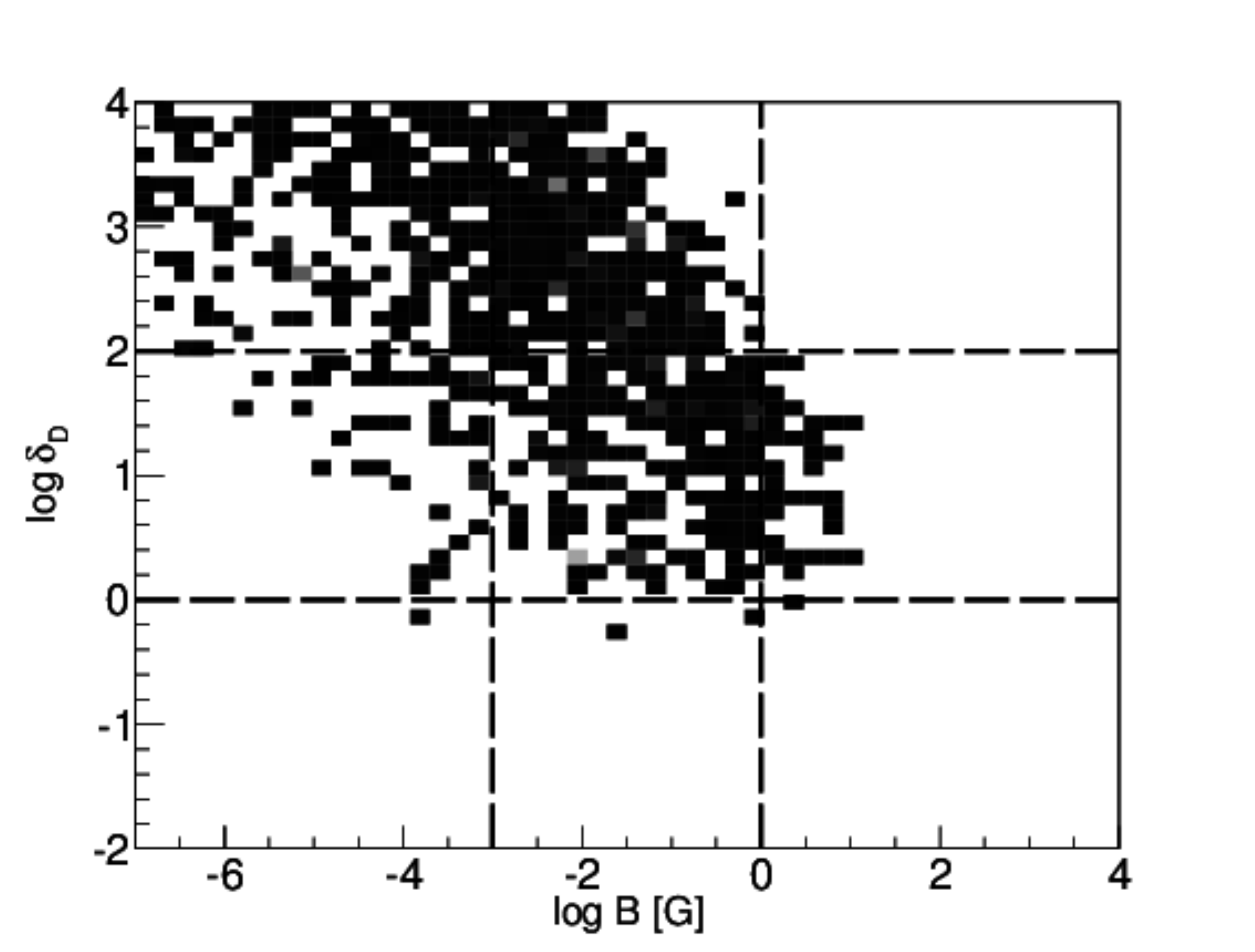}\\
   \includegraphics[clip,width=8.0cm]{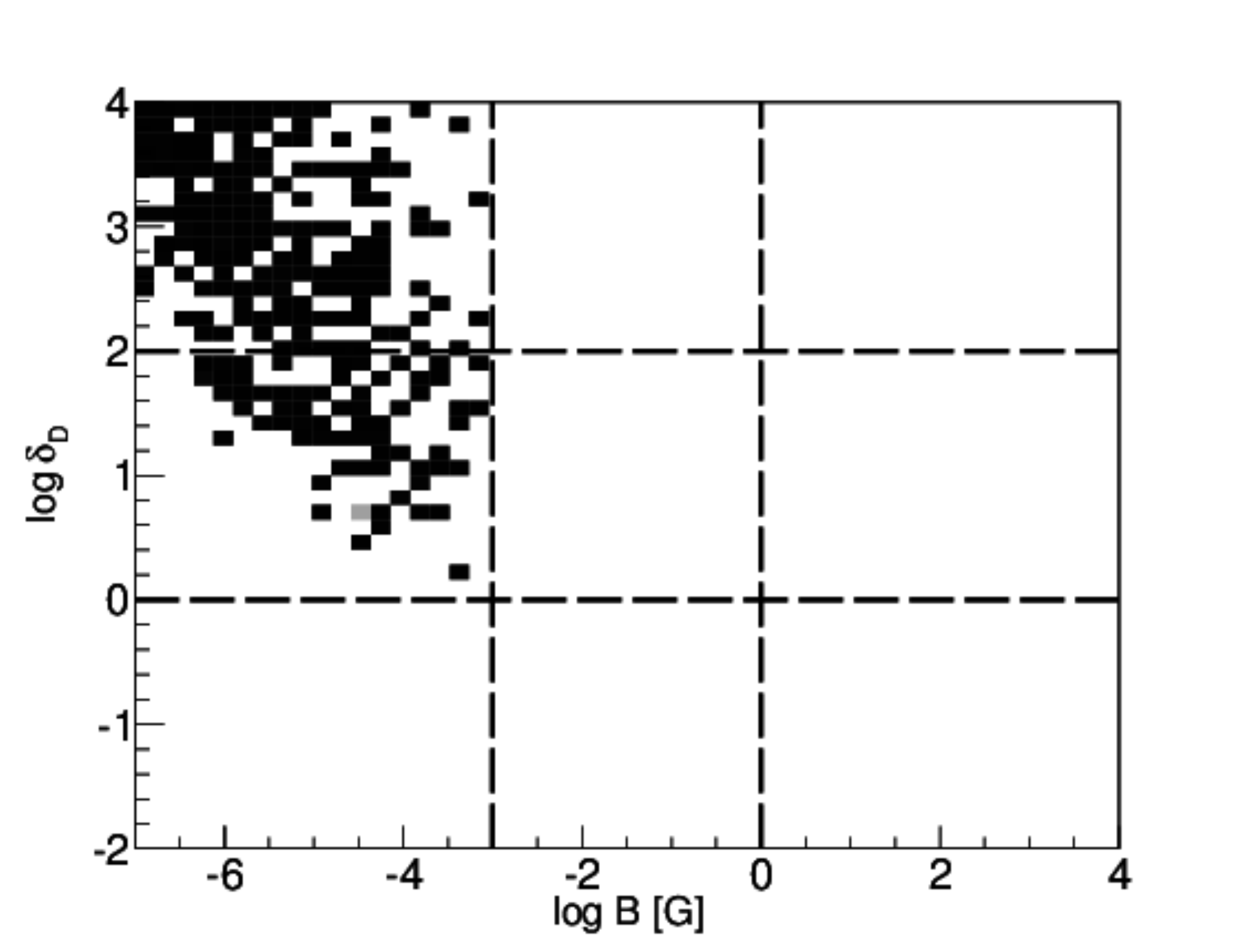}
   \end{center} 
    \caption{Posterior probability distributions on the
      $B$--$\delta_{\rm{D}}$ plane for the data on MJD 55598 (upper)
      and MJD 55623 (lower). The region bounded by the dashed lines
      indicate the reasonable ranges for blazars: $\log B = -3 $--$0$
      and $\log \delta_{\rm{D}} = 0 $--$2$.} \label{fig:soukan2} 
\end{figure}

The upper panel of Figure \ref{fig:soukan2} is the same as the top 
panel of Figure \ref{fig:notjizenhist}, but displayed for a narrower 
range of $\log B$ and $\log \delta_{\rm{D}}$. The dashed lines
indicate a reasonable region of the parameters which were reported in
past studies of the object: $B = 10^{-3} $--$ 1$ and
$\delta_{\rm{D}} = 1 $--$ 10^2$ (e.g., \cite{Tramacere2009};
\cite{Donnarumma2009}; \cite{Bartoli2016}). The posterior distribution
shown in the upper panel covers the reasonable region. The lower panel
shows that obtained with the SED data on MJD 55623.  In the lower
panel, a very small value of $B$ ($\log B$ $<$  $-4$) is required for
the optimal solution. Such an extraordinary parameter value implies
that a one-zone SSC model is not valid for the data. In our
SED analysis in Section 5, we do not include the data with which the
68.3\% confidence region of the posterior probability map obtained
with the non-informative prior does not include a region of
$\log B >-4$. The discarded data is those on MJD 55215, 55220, 55275,
55312, 55570, 55623, 55633, 56300, 56414 and 56428. We note
that very long $T$ ($10^{8-9}\;{\rm s}$) are required in the SED
analysis of those discarded data, although the X-ray flux shows
significant variations in days ($\sim 10^{5-6}\;{\rm s}$). It also
supports that the model is inappropriate for the data. We only
discuss 31 SED in the following sections, excluding those 10 samples.

\begin{table*}
\caption{Correlation coefficients between parameters} 
\begin{center}
  \begin{tabular}{c c c c c c c c c } \hline
 & $\log B$ & $\log \delta_{\rm{D}}$ & $\log T$ & $\log K_{\rm{e}}$ & $\log \gamma_{\rm{b}}$ & $p_0$&$p_1$  \\ \hline
$\log B$&1.000 &$-$0.996&0.993&0.995&$-$0.0744&$-$0.0920&$-$0.103\\
$\log \delta_{\rm{D}}$ & $-$0.996&1.000 &$-$0.999&$-$0.999&$-$0.00909&0.0394&0.0622\\
$\log T$&0.993&$-$0.999&1.000 &0.999&0.0468&$-$0.0155&$-$0.0437\\
$\log K_{\rm{e}}$&$-$0.995&$-$0.999&0.999&1.000 &0.0223&$-$0.0359&$-$0.0597\\
$\log \gamma_{\rm{b}}$ &$-$0.0744&$-$0.00909&0.0468&0.0223&1.000 &0.690&0.536\\
$p_0$&$-$0.0920&0.0394&$-$0.0155&$-$0.0359&0.690& 1.000&0.577\\
$p_1$&$-$0.103&0.0622&$-$0.0437&$-$0.0597&0.536&0.577&1.000 \\\hline
  \end{tabular}
\end{center}
\label{tab:soukan}
\end{table*}

\section{Time-series analysis}

As described in Section 3, we found strong correlations between
$\log B$, $\log \delta_{\rm{D}}$, $\log T$, and $\log K_{\rm{e}}$.
We need a prior distribution of only one of these parameters to
uniquely determine the optimal solution. We can assume a reasonable
prior distribution of $T$ from the light curve analysis. In this
section, we estimate the variation time-scale by modeling the
X-ray light curve of Mrk 421 with the Ornstein--Uhlenbeck (OU)
process. The OU process has been used to characterize the variations
observed in AGN, and also in blazars. It has been proposed that the
variability of AGNs has two characteristic time-scales
(\cite{Arevalo2006}; \cite{McHardy2007}; \cite{Kelly2011}). 
\citet{Sobolewska2014} reported that the $\gamma$-ray variability of
blazars can also be reproduced by a model with two time-scales, a
short time-scale of $\lesssim 1\;{\mathrm{d}}$ and a long time-scale
of $\gtrsim 100\;{\mathrm{d}}$. In this study, we use two kinds of
X-ray light curves of Mrk 421. The first was obtained from
high-cadence observations by ASCA for estimating the short
time-scale. The second was obtained from low-cadence observations by
$\it{Swift}$-XRT for the long time-scale. 

\subsection{Data}

We analyzed the data of Mrk 421 observed with ASCA \citep{Tanaka1994}
between MJD 50923 and 50934. We used the standard cleaned data. 
This satellite carries the Solid State Imaging Spectrometer (SIS;
\cite{Burke1994}; \cite{Yamashita1997}) and the Gas Imaging
Spectrometer (GIS; \cite{Ohashi1996}; \cite{Makishima1996}).
We used only the SIS0 detector operating in the 1-CCD mode. We binned
the light curve of Mrk 421 in intervals of 480 s, and showed it in the
upper panel of Figure \ref{fig:GPy}. 

We created the light curve of Mrk 421 observed with $\it{Swift}$-XRT
from November 2009 to April 2014 by the method described in Section 2.
The light curve is shown in the lower panel of Figure \ref{fig:GPy}.

\subsection{OU process}

We estimated the variation time-scale of the light curves by using the
OU process regression. The
model has three parameters: the variation time-scale, $\tau$, the
variation amplitude, $A_{\rm exp}$, and the Gaussian noise variance,
$\sigma^2$. We consider $N$ time-series data as a sample from a 
$N$-dimensional normal distribution, $\mathcal{N}(\bm{0},S)$, where
$S$ is an $N \times N$ variance--covariance matrix. The OU process is
defined with $S$ as follows: 
\begin{equation}\label{eq:6}
S_{ij}=A_{\rm{exp}} \exp  (- \frac{|t_i - t_j|}{\tau}) 
\end{equation} 
where $t_i - t_j = \Delta t$  is the time interval between the
$i^{\rm th}$ and $j^{\rm th}$ epochs \citep{Uhlenbeck1930}.
We consider the observed time-series data, $m(t)$, and a sample from
the OU process, $f(t)$. The data is obtained as $m(t) = f(t) +
\varepsilon$, where $\varepsilon$ is the error term following the
normal distribution, $\mathcal{N}(0,\sigma^2)$.

The OU process can be considered a special case of the Gaussian
process (GP) with an exponential kernel. In this paper, we used the
python framework, {\tt GPy}, to estimate the posterior probability
distributions of three parameters, $\tau$, $A_{\rm{exp}}$, and
$\sigma^2$. We assumed a constant $\sigma^2$ for all light
curve data which has different measurement error values. Mrk~421 is so
bright in X-rays that the error of the X-ray flux is small even in the
faint state. The typical fractional error is 0.02--0.04 in the X-ray
faint state and less than 0.02 in the X-ray bright state, while the
fractional variation amplitude is $\sim 10$. We neglected the
small difference in the errors for the OU process regression. The
posterior distributions were estimated with a hybrid Monte Carlo
method implemented in {\tt GPy}.

\subsection{Estimations of variation time-scales}

\begin{figure}[t]
 \begin{center}
\centering
\subfigure{\includegraphics[width=7cm]{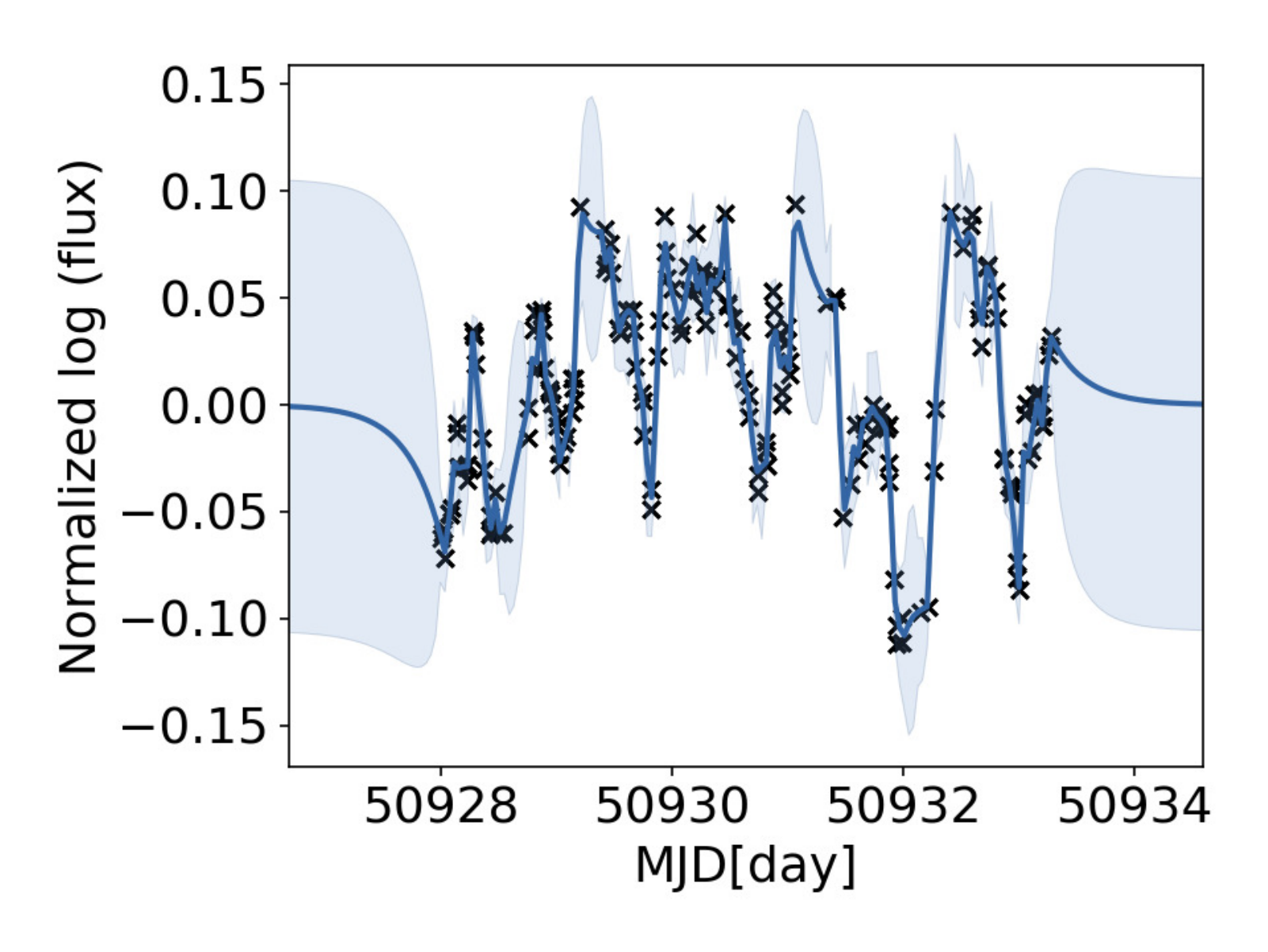}}\\
\subfigure{\includegraphics[width=7cm]{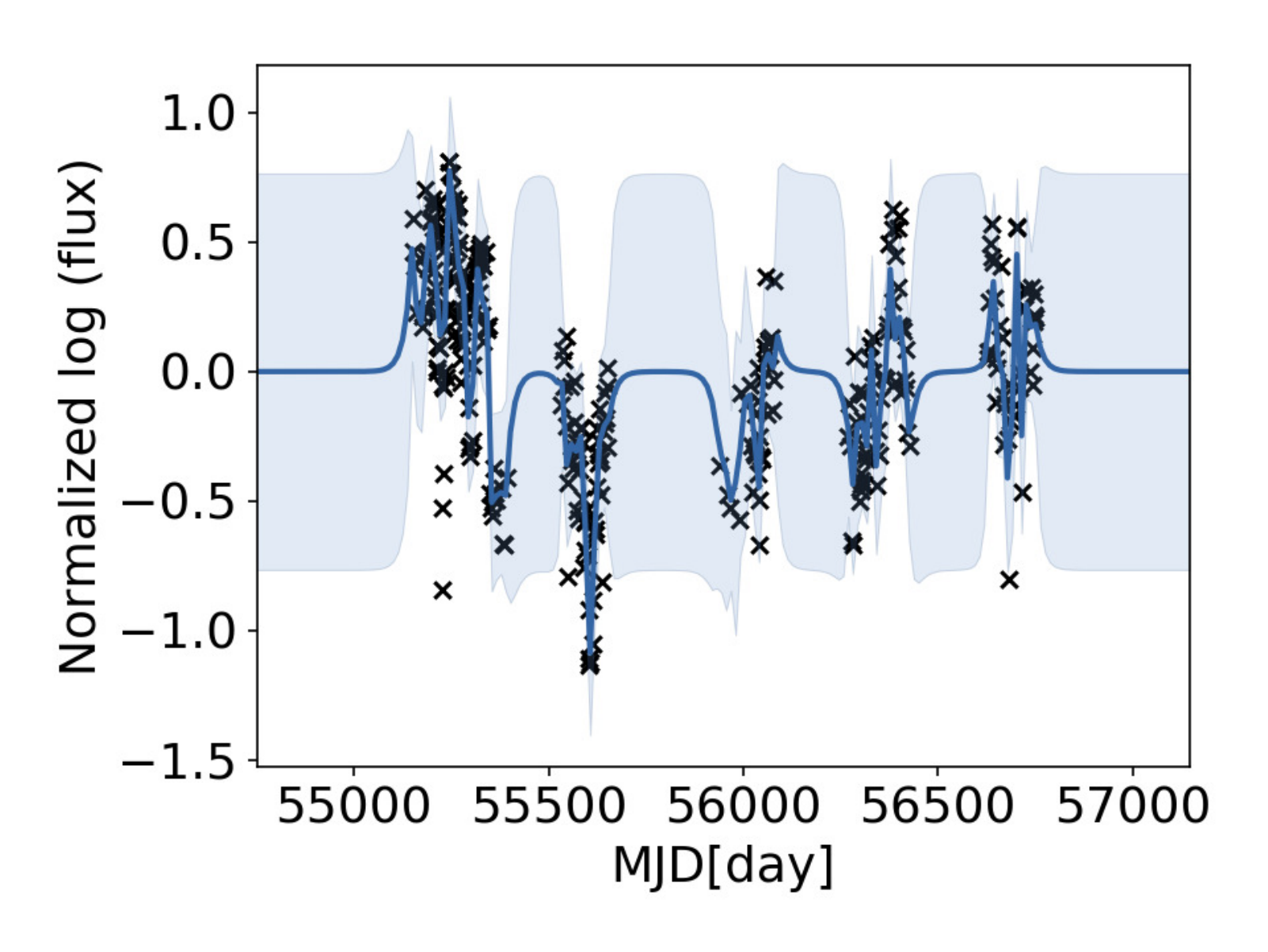}}
\end{center}
\caption{Data (crosses) and best-fit models (blue lines) of the OU
  process. The upper and lower panels show the ASCA and
  $\it{Swift}$-XRT data, respectively. The shaded region indicates the
  95\% confidence interval at each epoch.}\label{fig:GPy}
\end{figure}

\begin{figure}
\begin{center}
\centering
\subfigure{\includegraphics[width=7cm]{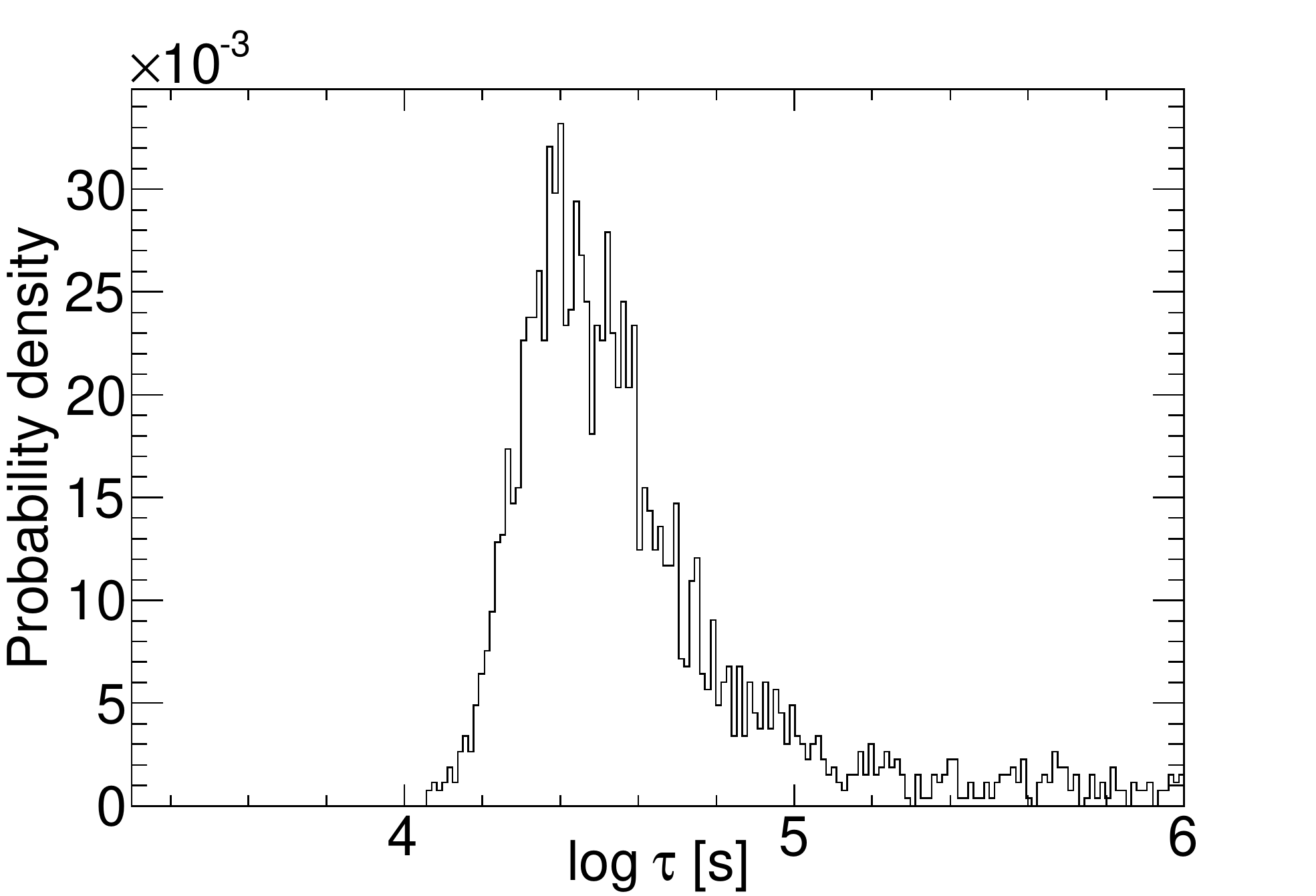}
 \label{fig:fig01left2}}
\subfigure{\includegraphics[width=7cm]{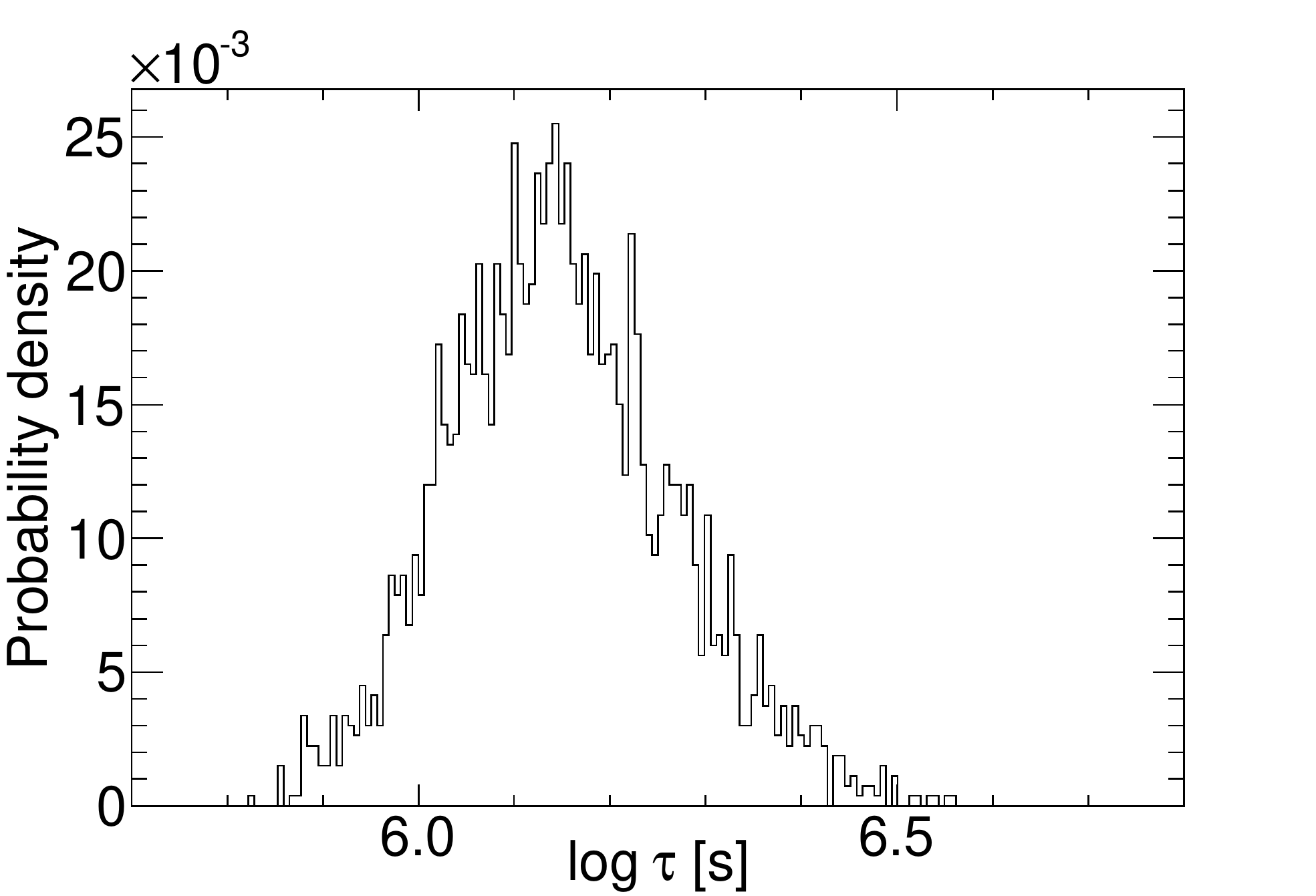}
 \label{fig:fig01right2}}
\end{center}
\caption{Posterior probability distribution of the time scale $\tau$
  plotted on a logarithmic scale. The distribution was
  obtained from the MCMC samples shown in Figure
  \ref{fig:Ttrace}.}\label{fig:Thist}
\end{figure}

We optimized the three parameters of the OU process for each light
curve, that is, the ASCA and $\it{Swift}$-XRT data. The light curves
and best-fit models are shown in Figure \ref{fig:GPy}.
Figure \ref{fig:Thist} shows the posterior distributions of
$\tau$. The optimal parameters of the OU process are listed in 
Table \ref{tab:GPy}. The means and standard deviations of the MCMC
samples of $\log\tau$ are $\mu = 4.5$ and $\sigma = 0.2$ for the ASCA data
and $\mu = 6.1$ and $\sigma = 0.1$ for the $\it{Swift}$-XRT
data.

We used them for the prior distributions of $T$ in our SED analysis,
assuming that $\tau$ corresponds to the light crossing time of the
emitting region. We note that the true $T$ could be shorter than
$\tau$ if the observed variations are governed by time-scales longer
than the light crossing time. It is unclear whether the short-term
variations are present in all states or only in a specific state of
Mrk 421. In this paper, we use the short time-scale estimated from the
ASCA data as the shortest $T$. Since the optimal solution of the SED
model depends on the assumed parameters, it is important to see the
result with both the minimum and maximum $T$.

\begin{table}
\caption{Optimal parameters of OU process.} 
\begin{center} 
\begin{tabular}{c c c c c} \hline
&ASCA &$\it{Swift}$/XRT &  \\ \hline
$\log \tau$ [s]&$4.4^{+0.3}_{-0.2}$ &$6.1^{+0.1}_{-0.1}$ & \\
$A_{\rm{exp}}$&$2.9^{+0.4}_{-0.3}\times 10^{-3}$ &$1.4^{+2}_{-2}\times 10^{-1}$ & \\
$\sigma^2$&$4^{+3}_{-2}\times 10^{-5}$ &$8^{+3}_{-3} \times 10^{-3}$ &\\\hline
 \end{tabular}
\end{center}
\label{tab:GPy}
\end{table}

\section{Results of the SED analysis}
\subsection{Results obtained with the prior distribution of $T$}

\begin{figure}
  \begin{center}
    \includegraphics[clip,width=8cm]{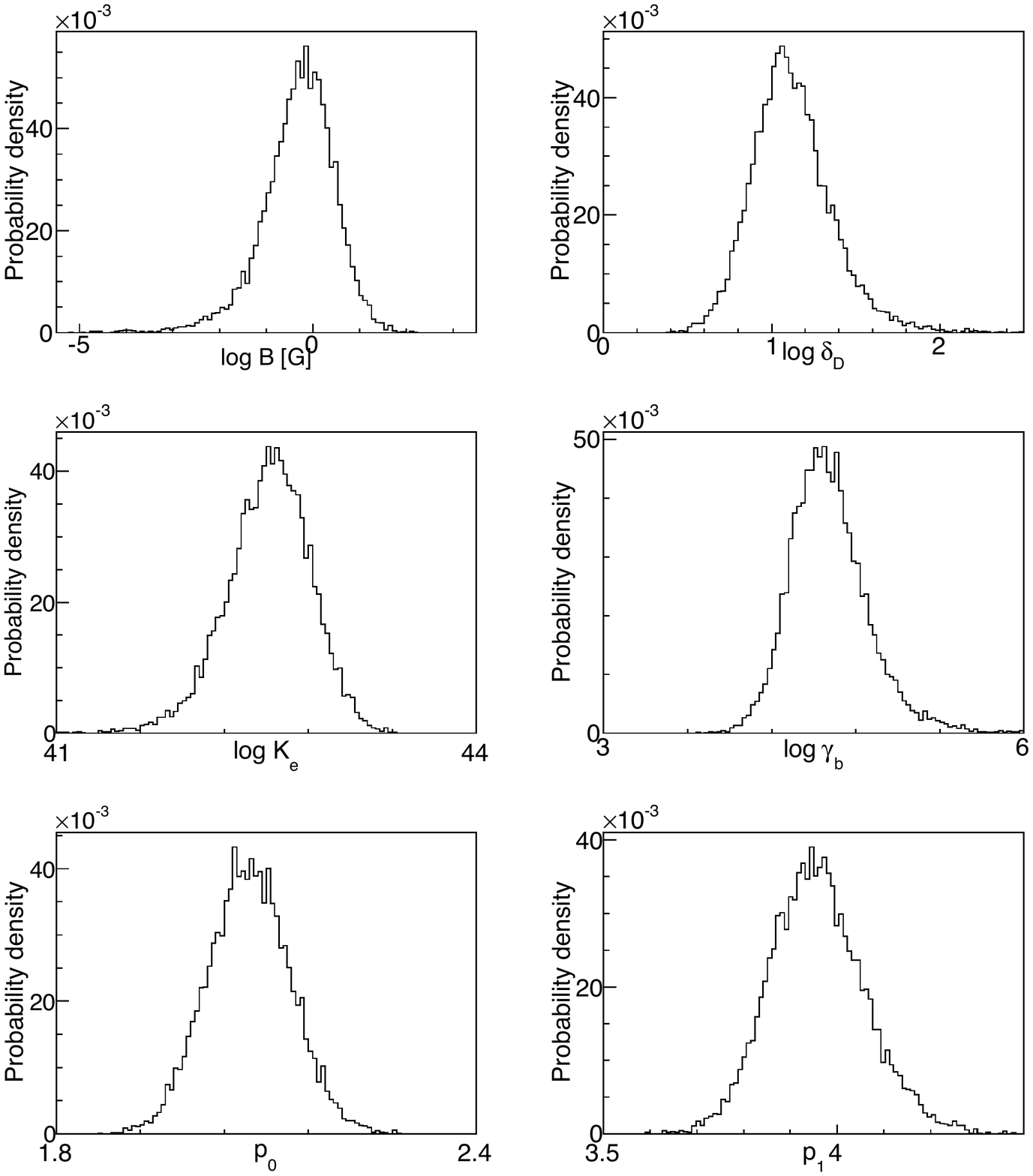}
\end{center}
  \caption{ Probability distributions of six parameters, $\log B$,
    $\log \delta_{\rm{D}}$, $\log K_{\rm{e}}$, $\log \gamma_{\rm{b}}$,
    $p_0$, and $p_1$.}\label{fig:T0.3hist}
\end{figure}

We estimated the model parameters of the SED of 31 epochs
from 2009 to 2014, using the prior distribution of $T$
estimated in Section 4.

First, we show the results obtained with the prior for the short 
time-scale estimated from the ASCA data. The prior of $\log T$ was set
to be a normal distribution with a mean of $\log T=4.5$ and a
standard deviation of 0.2. Figure \ref{fig:T0.3hist} shows their
posterior distributions obtained with the data on MJD 55272, as an
example. We can uniquely obtain the optimal solution and their
uncertainties. The distributions are almost symmetric. In this
paper, the optimal parameters are the means of each set of the MCMC
samples, and the uncertainty of the optimal parameters is the
68.3$\%$ confidence interval of them.

\begin{figure*}
  \begin{center}
    \includegraphics[width=160mm]{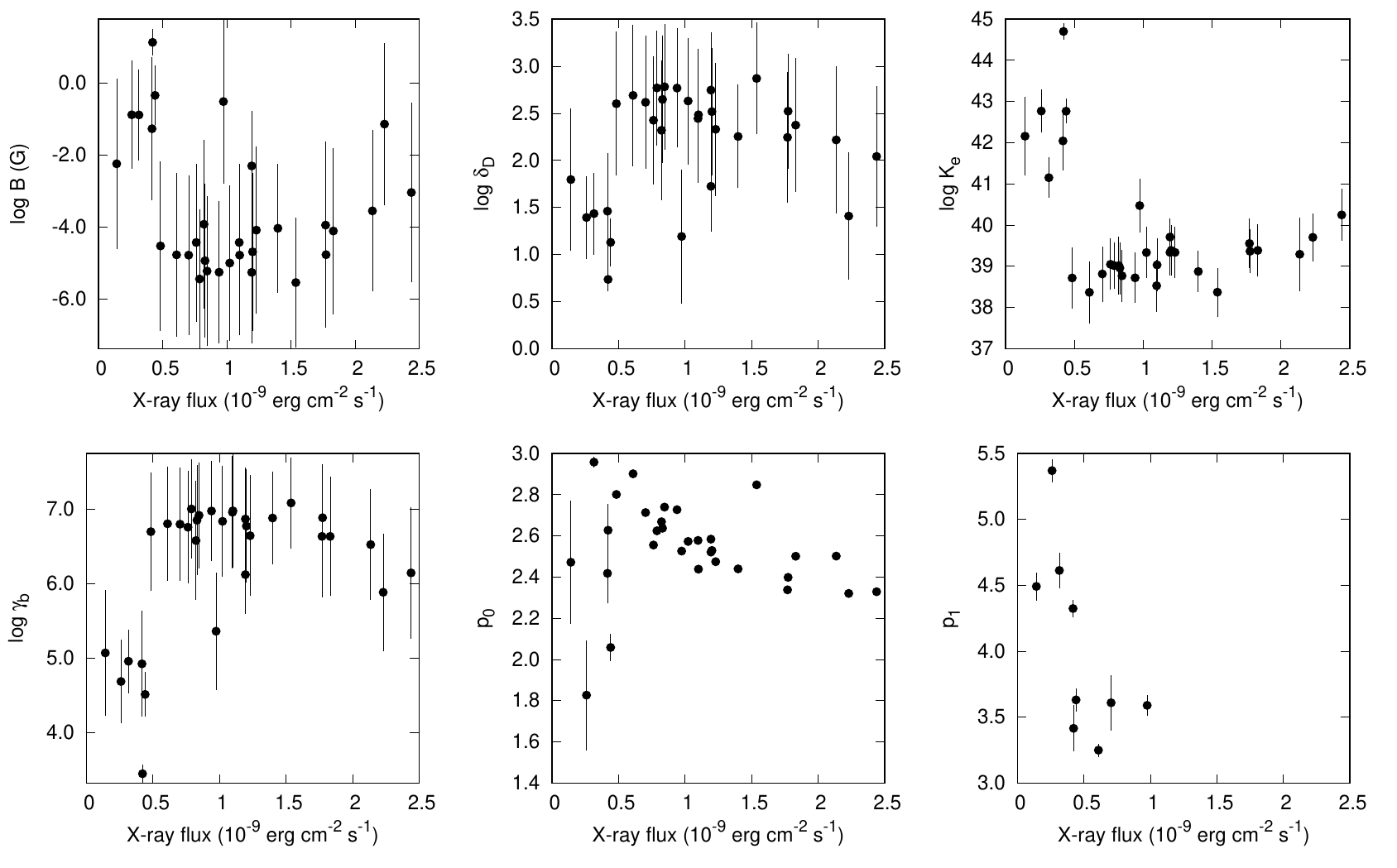}
  \end{center}
  \caption{Correlation of X-ray flux and the values of the parameters
   $\log B$, $\log \delta_{\rm{D}}$, $\log K_{\rm{e}}$,
   $\log \gamma_{\rm{b}}$, $p_0$, $p_1$, for the model with a prior
   distribution of $\log T = 4.5 \pm 0.2$. In the panel of $p_1$, we
   have removed the data in which $p_1$ was fixed to $20$. The errors
   of the X-ray flux are smaller than the symbol size.}\label{fig:T0.3correlation}
\end{figure*}

Figure \ref{fig:T0.3correlation} shows the correlation of the X-ray
flux and the parameter values obtained with the short $T$ prior.
As can be seen in this figure, $\delta_{\rm D}$ is extraordinary
large ($>100$) in the X-ray bright state. $\delta_{\rm D}$
becomes further large if the observed variation time-scale is longer
than the light crossing time, because $\delta_{\rm D}$ is
anti-correlated with $T$, as shown in Figure~\ref{fig:notjizenhist}.
In the X-ray faint states, the mean of $\delta_{\rm{D}}$ is 24.0
($\log \delta_{\rm D} = 1.38$), and this value is typical of the
Doppler factor of Mrk~421 (e.g., \cite{Abdo2011}; \cite{Bartoli2016}). 

\begin{figure*}
  \begin{center}
    \includegraphics[width=160mm]{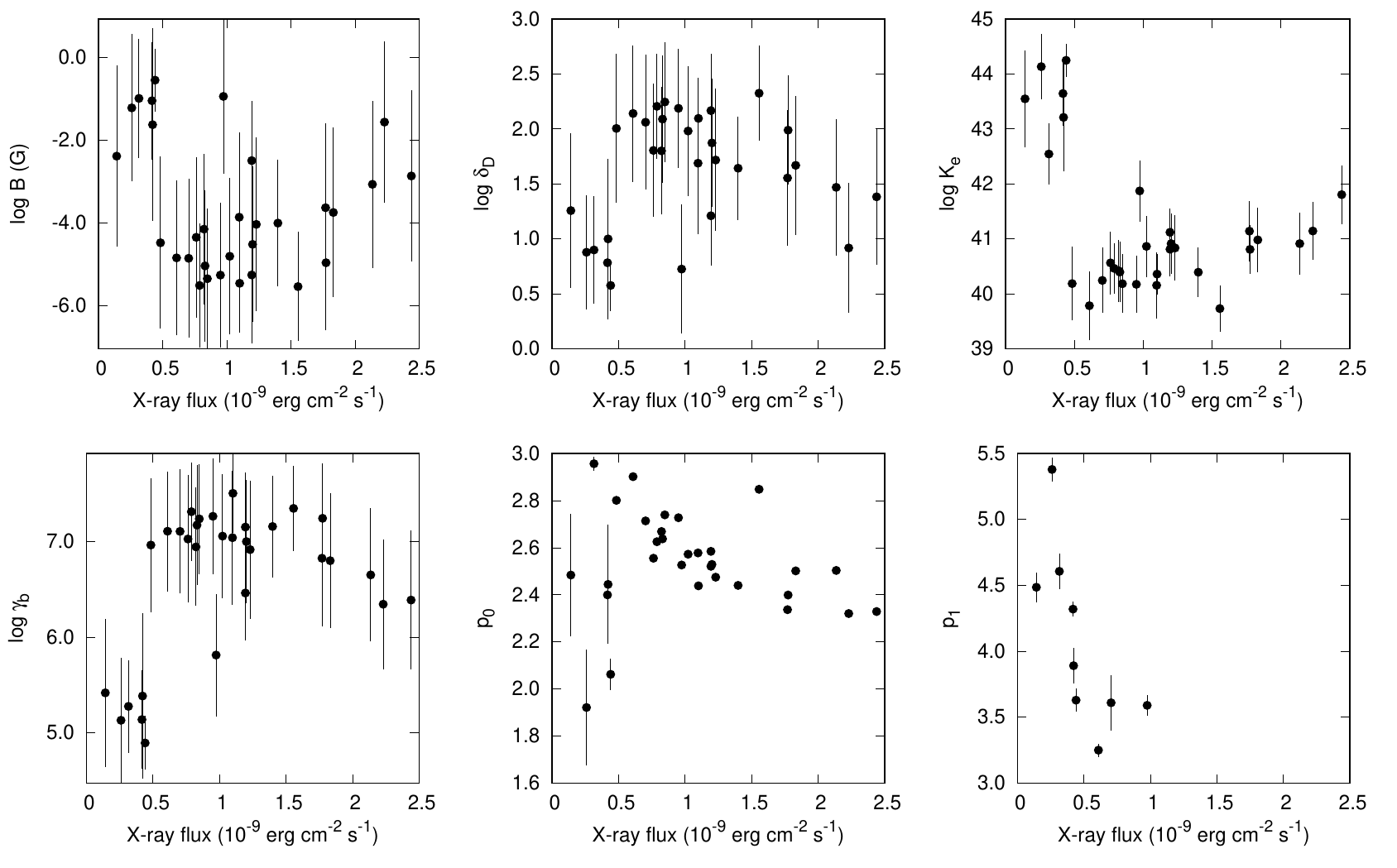}
  \end{center}
  \caption{As Figure~\ref{fig:T0.3correlation} but for the model with a
   prior distribution of $\log T = 6.1 \pm 0.1$.}\label{fig:T15correlation}
\end{figure*}

Second, we show the results obtained with the long $T$ prior. 
The prior of $\log T$ was a normal distribution with a mean of
$\log T=6.1$ and a standard deviation of $0.1$.
Figure~\ref{fig:T15correlation} shows the results in the same way as
Figure~\ref{fig:T0.3correlation}, but for the model with the long $T$
prior. We found that most of the bright state has acceptable
$\delta_{\rm D}$ ($< 100$), while the faint state tends to have
atypically small $\delta_{\rm D}$ ($< 10$) for SED studies of the
object. 
 
The results imply that the time-scale or the size of the dominant
emitting region changed data-by-data. In other words, a model with a
common prior distribution of $T$ is possibly unsuitable for the
observed SED. Then, we need to restrict one of $B$, $\delta_{\rm D}$,
and $K_{\rm e}$ in order to uniquely determine the solution for the
SED analysis, as mentioned in Section 3.

\subsection{Results obtained with the prior distribution of $\delta_{\rm D}$}

In this section we report on a SED analysis using a model with a fixed 
$\delta_{\rm D}$ instead of the $T$ prior.
\citet{Bartoli2016} estimated $\delta_{\rm D}$ of Mrk 421
to be 10--41, based on an analysis of the SED data with a broad
frequency coverage. This range of $\delta_{\rm D}$ is consistent with
those found in previous investigations (\cite{Abdo2011};
\cite{Shukla2012}). Here, we fixed $\delta_{\rm D}$ to be 20. We
confirmed that the MCMC converged to a stationary distribution and
that the posterior distributions of all parameters had a single peak,
which enable us to determine the optimal solution and their
uncertainties. 

\begin{figure*}
  \begin{center}
    \includegraphics[width=160mm]{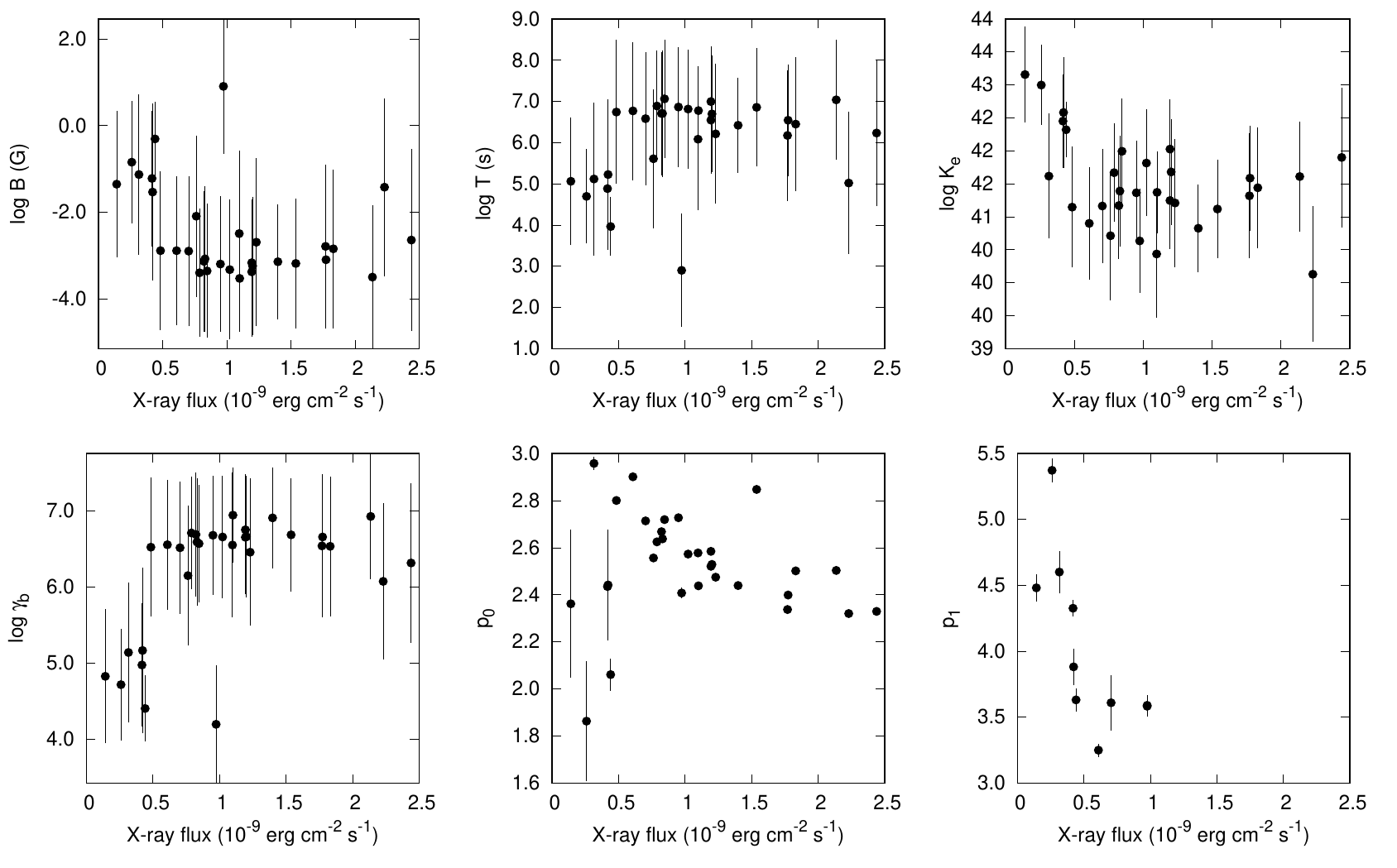}
  \end{center}
  \caption{As Figure \ref{fig:T0.3correlation} but for the model with
    the Doppler factor fixed to $\delta_{\rm D} = 20$.}\label{fig:Dcorrelation}
\end{figure*}

Figure \ref{fig:Dcorrelation} shows the correlation of the X-ray flux 
and the parameter values. The X-ray bright state shows small $B$ and
$K_{\rm e}$ and large $T$ and $\gamma_{\rm b}$ compared with the faint
state. In the X-ray bright state, $p_0$ exhibits an anti-correlation
with the X-ray flux. $p_0$ remains relatively small in the faint
state. No clear correlation can be seen between $p_1$ and X-ray
flux. We note that in the panel for $p_1$, we removed the most samples
of the bright state because $p_1$ of those samples was set to be 20,
as mentioned in Section 3.1. These trends between the parameters and
X-ray flux can be seen also in Figures \ref{fig:T0.3correlation} and
\ref{fig:T15correlation}. Thus, they are independent of the priors
used in this analysis.

\begin{figure}
 \begin{center}
  \includegraphics[width=8.0cm]{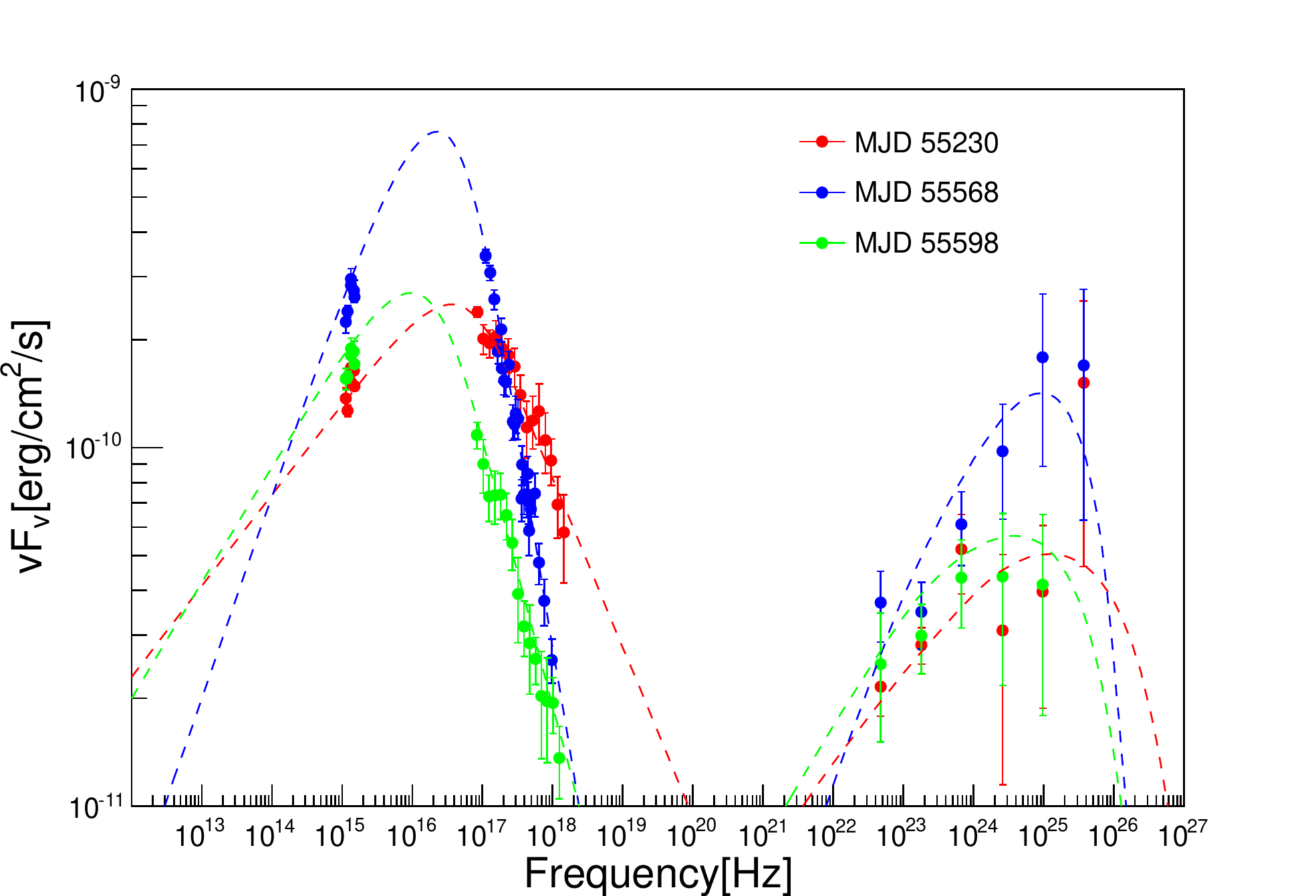}
  \includegraphics[width=8.0cm]{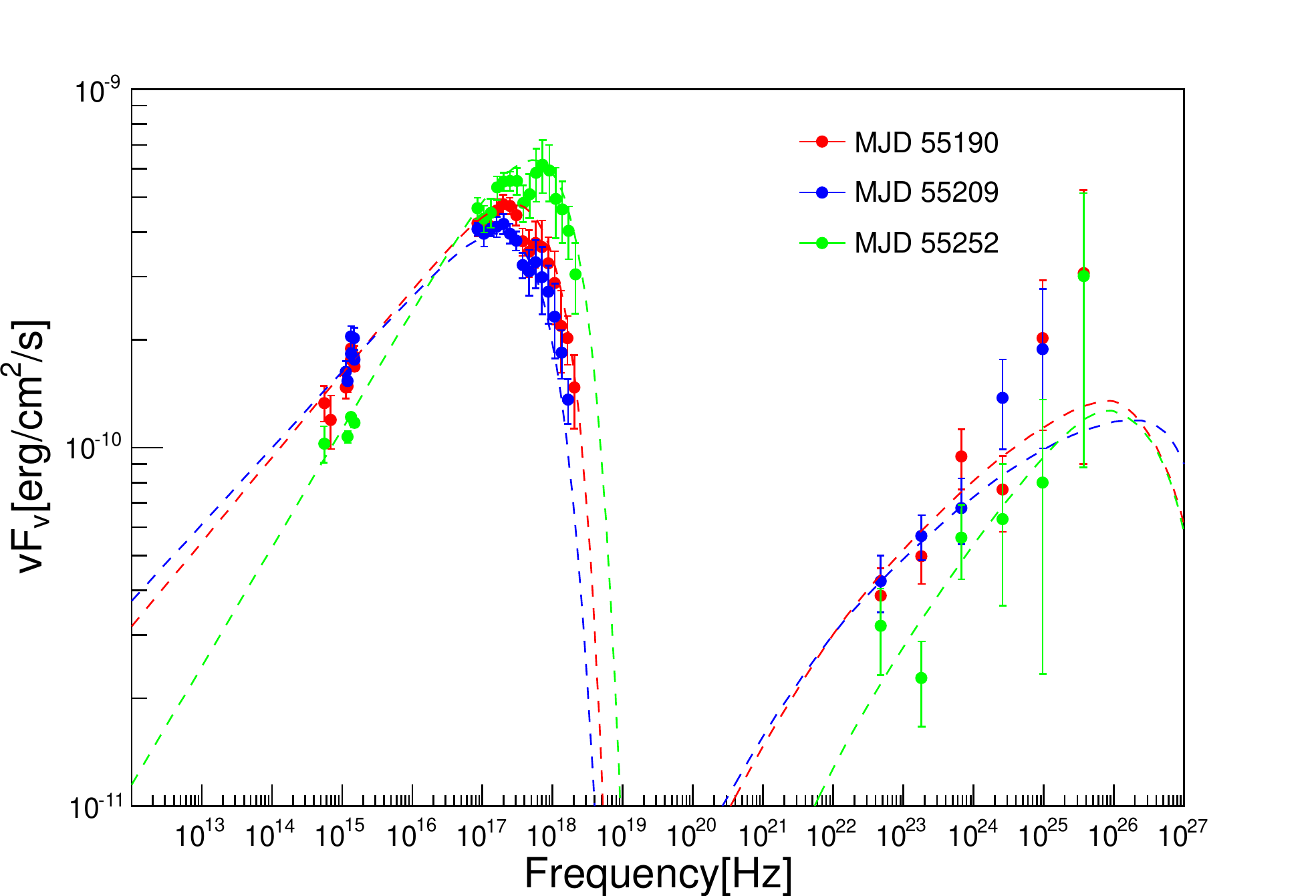}
\end{center}
    \caption{Examples of observed and model SEDs of Mrk 421. The model
      SED are indicated by the dashed lines. The left and right
      panels show examples in the faint and bright
      state, respectively. The epoch of the data is indicated in each
      panel.}\label{fig:sed} 
\end{figure}

Examples of the observed and modeled SED are shown in
Figure~\ref{fig:sed}. The top and bottom panels show examples in the
faint state and the bright state. We can confirm that the one-zone SSC
model well reproduces the data for the set of SED considered
in this section (see Section 3.3 about the discarded
samples). The reduced $\chi^2$ for the SED model optimization are
shown in Appendix 1.

\section{Discussion}
\subsection{General properties of SSC modeling of blazar SED}

It is widely known that, in most cases, observations cannot
constrain all SSC parameters because of the strong correlations
between a part of them (e.g. \cite{Tavecchio1998}). To date, the detailed
structure of the correlations has not explicitly shown for the
optimization problem of the SSC model. Figure~\ref{fig:notjizenhist}
successfully provides it in the form of the posterior probability
distribution. We can obtain important properties of SSC modeling of
the blazar SED from Figure \ref{fig:notjizenhist} and results in
section 3.3.

First, the optimization of the model should be performed on a
logarithmic scale for $B$, $\delta_{\rm{D}}$, $T$, $K_{\rm{e}}$, and
$\gamma_{\rm{b}}$. We found that $\log B$, $\log \delta_{\rm{D}}$,
$\log T$, and $\log K_{\rm{e}}$ have a linear correlation in Figure
\ref{fig:notjizenhist}, which can be readily sampled with the
Metropolis algorithm. Several past studies using MCMC estimated the
optimal parameters without TeV $\gamma$-ray data or an informative
prior distribution (e.g., \cite{Dahai2013}; \cite{Ding2017}). We
should note that non-optimal solutions can be obtained when the parameter
space is insufficiently explored with MCMC. 

Second, care should be taken when using priors for two or more
parameters. As shown in Figure \ref{fig:notjizenhist}, a prior
distribution of one of the four parameters gives constraints on the
other three parameters. We have confirmed that the solution was
uniquely determined when a Gaussian prior was set for $B$ or
$K_{\rm e}$, instead of $T$ or $\delta_{\rm D}$. In past studies, the
SSC model has sometimes been optimized with both fixed
$\delta_{\rm D}$ and $T$ (e.g., \cite{Tramacere2009};
\cite{itoh2015}). Such multiple priors may lead to the optimal model
parameters being overlooked. For example, in the bottom panel of
Figure \ref{fig:soukan2}, the solution obtained with conventional
constraints on the parameters (the region surrounded by the dashed
lines in the figure) is different from the original optimal solution
without any constraints. In such a case, we should reconsider the
validity of the constraints or the model itself. Therefore, it is
important to first carefully examine the posterior probability
distribution in the parameter space without informative priors.

If one of $B$, $\delta_{\rm{D}}$, $T$, and $K_{\rm{e}}$ is constrained
by a prior distribution or fixed value, the other parameters are
determined. In other words, the values of the other parameters depend
on the constraint. Care should be taken in discussions based
on the values estimated by a given constraint. However, we
confirmed the common trend in the parameters regardless of the fixed
value or prior distribution. For example, the results in Figures
\ref{fig:T0.3correlation} and \ref{fig:T15correlation} differ in the
values of $B$, $\delta_{\rm{D}}$, and $K_{\rm{e}}$ because the centers
of the prior distributions of $T$ are different, though the features
of the relative variations of the parameters is common between
them. Therefore, we can discuss the characteristic trend of the
parameters regardless of the constraint.

Constraints are employed for $T$ and $\delta_{\rm D}$ in the study of
SSC modeling of blazar SED because they can be estimated from other
observations, for example, the temporal variability of the flux or
that of VLBI images. As mentioned in Section 5.1, $T$ or the size of
the emitting region of the bright state is distinct from that of the
faint state. In Section 5.2 we reported on the results with the prior
distribution of $\delta_{\rm D}$. We assumed a constant
$\delta_{\rm D}$ for all epochs. However, there is a possibility that
$\delta_{\rm D}$ also changed with time. In this paper, we only
discuss the features of the parameter variations which are common for
both the results of Section 5.1 and 5.2. 

\subsection{Physical conditions of the jet}

As described in Section 5.2, $\gamma_{\rm{b}}$ tends to be large in
the X-ray bright state compared with the faint state. 
$\gamma_{\rm{b}}$ is possibly interpreted as a cooling break
(\cite{Kino2002}). However, it is unlikely for our result because
$p_1$ is estimated to be larger than that expected from this scenario 
($p_1-p_0=1$), as reported in \citet{Bartoli2016}: $p_1>5$ in most
samples of the X-ray bright state and $p_1-p_0>1.5$ even in the X-ray
faint state. In this case, we can consider that $\gamma_{\rm{b}}$ is
effectively the maximum energy that is achieved in the
acceleration process. Hence, the result indicates that the maximum
energy of electrons increased in the X-ray bright state. A scenario
with high-energy injection by internal shocks in relativistic shells
(the shock-in-jet model) is one way to explain such variability (e.g.,
\cite{Spada2001}). As well as the shock-in-jet scenario, the variation
in $\delta_{\rm D}$  has also been proposed as a possible origin of
the observed flux  variations. Figure \ref{fig:T0.3correlation} and
\ref{fig:T15correlation} possibly support that $\delta_{\rm D}$
actually increased in the X-ray bright state. However, we emphasize
that the increase in $\gamma_{\rm b}$ is seen also in Figures
\ref{fig:T0.3correlation} and \ref{fig:T15correlation}. Hence, this
feature of $\gamma_{\rm b}$ is independent of the model and prior.

\citet{itoh2015} and \citet{Bartoli2016} also reported that
$\gamma_{\rm{b}}$ increased during bright X-ray flares of Mrk~421,
which is consistent with our result. On the other hand,
\citet{itoh2015} described that small flares in the X-ray faint state
were due to the increase in $K_{\rm e}$. However, as can be seen from
Figures \ref{fig:T0.3correlation}, \ref{fig:T15correlation}, and
\ref{fig:Dcorrelation}, there is no positive correlation between
$K_{\rm e}$ and the X-ray flux in the X-ray faint state. 

In the bright state, the X-ray flux shows no clear correlation
with the SED model parameters, except for $p_0$. It decreases with
increasing X-ray flux, as can be seen all in Figures
\ref{fig:T0.3correlation}, \ref{fig:T15correlation}, and
\ref{fig:Dcorrelation}. This is because $p_0$ determines the ratio
between X-ray and optical fluxes, and the optical variations are
fractionally much smaller than the X-ray one.

\begin{figure}
 \begin{center}
       \includegraphics[width=8.0cm]{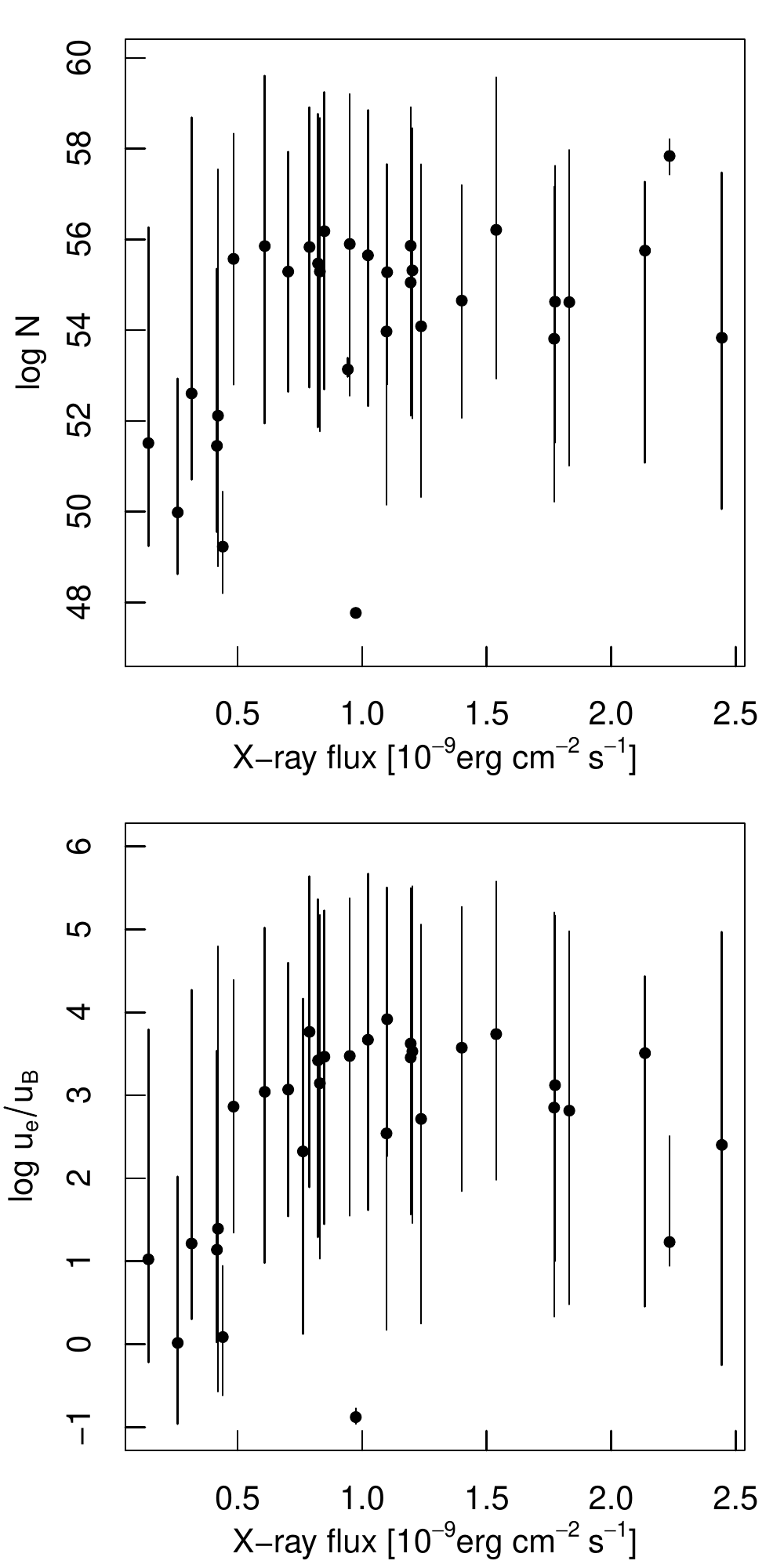}
\end{center}
 \caption{The total number of electrons (upper panel) and
   $u_e/u_B$ (lower panel) calculated from the results obtained in the
   $\delta_{\rm D}$-fixed model. The horizontal axis shows the X-ray
   flux.}\label{fig:xueub}
\end{figure}

The upper panel of Figure~\ref{fig:xueub} shows the total number of
electrons and the X-ray flux. The total number of electrons was
obtained with integrating $N_{\rm e}(\gamma)$ over $\gamma$ between
$\gamma_{\rm min}$ and $\gamma_{\rm max}$ based on the results 
obtained with the $\delta_{\rm D}$-fixed model. The bright state has
a larger number of electrons compared with the faint state.
We also calculated the ratio of the electron energy density
$u_e$ to the magnetic energy density $u_B$, and show it against 
the X-ray flux in the lower panel of Figure~\ref{fig:xueub}. The
electron energy is larger than the magnetic energy in all samples
($u_e/u_B \gtrsim 1$), except for the data on MJD~55319
($u_e/u_B = 0.13$). The energy ratio depends on $\gamma_{\rm min}$,
which we assumed $\gamma_{\rm min}=10^2$ in our analysis. The values
of the ratio become smaller by a factor of 0.1--0.8 with a larger
$\gamma_{\rm min}$ of $10^3$, while the ratio is greater than unity
even in this case.  The result of $u_e/u_B \gtrsim 1$ is consistent
with previous studies of this object (\cite{Abdo2011};
\cite{Bartoli2016}). The ratio is larger in the bright state than that
in the faint state. This trend of $u_e/u_B$ is independent of
$\gamma_{\rm min}$.

The SED analysis with $\delta_{\rm D}=20$ (Section 5.2) gives
$\log T = 4.7$ on average of the faint state. In contrast, the bright
state has a longer one: $\log T = 6.3$ on average. These values are
close to $\tau$ obtained from the time-series analysis with the OU
process (Section 4), that is, $\log\tau=4.4_{-0.2}^{+0.3}$ from the
high-cadence data and $\log\tau=6.1_{-0.1}^{+0.1}$ from the
low-cadence data, as shown in Table \ref{tab:GPy}. This agreement
supports the idea that the variation time-scales are actually
close to the light crossing time of the emitting region, and that
there are two emitting regions having different sizes. The larger
emitting region with the longer variation time-scale significantly
contributes to the X-ray flux in the bright state.

As mentioned in section 3.1, the size of the emitting
region $R$ is given with $R= c \delta_{\rm{D}} T/(1+z)$.
In this study, we fixed $\delta_{\rm{D}}$ = 20. However, it is
possible that $\delta_{\rm D}$ changed with time, which may make $R$
constant even if $T$ changes with time. However, this is unlikely in
the present study because $\delta_{\rm D}$ would have to have changed
by roughly two orders of magnitudes to cancel out the variation in $T$
between $\log T\sim 3$ and $7$. In the faint state, $R$ is calculated
to be $2.9 \times 10^{16}  \rm {cm}$ with $\log T = 4.7$. In the
bright state, $R$ is calculated to be $1.1 \times 10^{18}\rm{cm}$ with
$\log T = 6.3$. Past studies have reported that the optical emitting
region typically has $R=10^{15-16} \rm {cm}$ (e.g.,
\cite{Tramacere2009}, \cite{Abdo2011}). Therefore, the $R$ for the
faint state is reasonable for the current understanding of blazars,
while the $R$ for the bright state is quite large.

\citet{Spada2001} propose a model of the internal shocks between
plasma shells in the AGN jets. According to this model, the first
collision of the shells occurs in the upstream region of the jet at
$\sim 10^{16}\,{\rm cm}$ from the central black hole. The shells then
experience second and subsequent collisions in the downstream region
at $\sim 10^{17-20}\,{\rm cm}$. We propose that the radiation from
such a downstream region is responsible for the large emitting region
of the bright state. The emergence of an additional emitting region in
the 2010 outburst is also supported by the behavior of the optical
polarization reported in \citet{itoh2015}. The total number of
electrons in the bright state is larger than that of the faint
state. This is probably due to a large emitting area of the bright
state. The high $u_e/u_B$ of the bright state possibly
suggests that the magnetic energy is mostly transferred into the
electrons in the downstream region.

On the other hand, $p_0$ obtained by our analysis is larger than what
is expected by the first-order Fermi acceleration in shocks 
($p_0=2.0$--$2.2$) in the X-ray bright state. This discrepancy may
suggest a scenario where the observed emission is not SSC from a
single source (one-zone), but a combination of multiple SSC
components (e.g., \cite{Blazejowski2005}). The two distinct variation
time-scales possibly support this scenario. As mentioned in section
3.3, we discarded 10 SED samples because the optimal $B$ and $T$ are
beyond the reasonable ranges. This fact also implies that multiple SSC 
sources are required for a significant part of the epochs.

\begin{figure}
  \begin{center}
    \includegraphics[width=8cm]{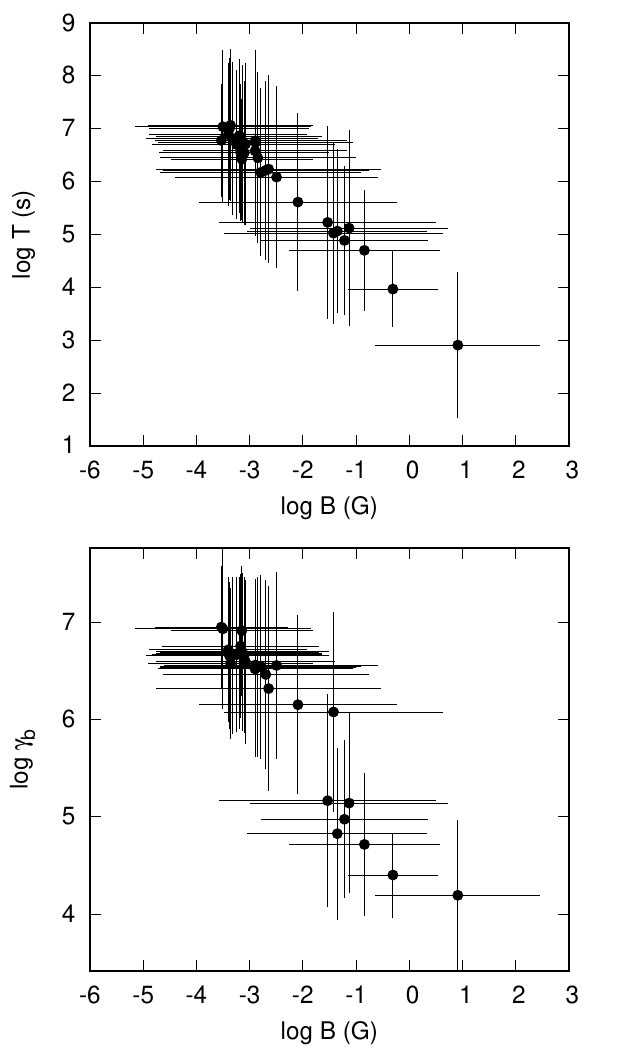}
  \end{center}
\caption{Correlation between the three parameters $B$, $T$, and
  $\gamma_{\rm{b}}$. The upper panel shows the scatter plot of $B$ and
  $T$. The lower panel shows that of $B$ and
  $\gamma_{\rm{b}}$.}\label{fig:parasoukan} 
\end{figure}

We found notable correlations between the estimated parameters of the
SSC model. Figure \ref{fig:parasoukan} shows the correlation between
$B$, $\gamma_{\rm b}$, and $T$ obtained from the
$\delta_{\rm D}$-fixed model in section 5.2. We confirmed that no
other combinations of parameters exhibit such clear correlations. The
negative correlation between $T$ and $B$ is reminiscent of the
relation between the synchrotron cooling time-scale and magnetic
field. The synchrotron cooling time-scale $t_c$ of an electron in a
homogeneous magnetic field in the observer's frame, can be estimated
as follows: 
\begin{equation}\label{eq:9}
t_c = (\delta_{\rm D})^{-1} 5\times 10^{11}
 (1+z)^{1/2}B^{-3/2}[\rm{G}](\nu_{\rm obs}[\rm{Hz}]/\delta_{\rm D})^{-1/2} [s]
\end{equation}
where $\nu_{\rm obs}$ is the observation frequency \citep{Tucker1975}.
The cooling time-scale decreases with a larger $B$ at a given                  
$\nu_{\rm obs}$. The shortening of the cooling time-scale in the X-ray
regime may result in a decrease in the number of electrons emitting
X-rays, and thereby may be responsible for the decrease in
$\gamma_{\rm b}$. Hence, the correlation seen in the lower panel of
Figure \ref{fig:parasoukan} can be explained by this
scenario. However, equation~(\ref{eq:9}) predicts the slope of the
$\log T$--$\log B$ relation to be $-3/2$, while it is $\sim -1.0$ in
the top panel of Figure~\ref{fig:parasoukan}. 

\section{Summary}
In this paper, we estimated the SSC model parameters for the SED of
the blazar Mrk 421 using the Markov chain Monte Carlo (MCMC)
method. We used the optical--UV (UVOT/$\it{Swift}$ and Kanata), X-ray
(XRT/$\it{Swift}$), and $\gamma$-ray ($\it{Fermi}$-LAT) data sets
obtained between 2009 and 2014. We found a strong correlation between
the four SSC model parameters, that is, $\log B$,
$\log \delta_{\rm D}$, $\log T$, and $\log K_{\rm e}$ for a model with
non-informative priors. The correlation makes the model degenerate. As
a result, the optimal solution can be determined uniquely only when
one of the four parameters is restricted. We used two models, a model
with a prior on $T$ and one with a prior on
$\delta_{\rm D}$. Using a narrow prior distribution for $T$
results in unphysically large or small values for $\delta_{\rm D}$
depending on the X-ray flux. The
results suggest that $T$ was large in the X-ray bright state compared
with the faint state. Our SED analysis gives $\log T = 6.3$ in the
X-ray bright state and $\log T = 4.7$ in the faint state, which are
consistent to the characteristic time-scales obtained from the X-ray
light curves. Those short and long $T$ correspond to the sizes of the
emitting region of $2.9\times 10^{16}  \rm {cm}$ and
$1.1\times 10^{18}  \rm {cm}$, respectively. The X-ray flares are due
to an increase in $\gamma_{\rm b}$. The large emitting region in the
bright state is possibly originated from the downstream area in the
jet where the magnetic energy is mostly converted to electrons.
On the other hand, the presence of the two kinds of emitting
areas implies that the one-zone model is unsuitable to reproduce, at
least a  part of the observed SEDs.

The MCMC code used in this paper is available on request.

\begin{ack}
The authors thank the {\it Suzaku}, {\it Swift}, and {\it Fermi} teams
for the operation, calibration, and data processing. Y. ~F. was
supported by JSPS KAKENHI Grant Numbers 2400000401 and
2424401400. M. ~U. was supported by JSPS KAKENHI Grant Number
25120007. We would like to thank Ioannis Liodakis and Shiro Ikeda for
their Valuable comments on this research. We also thank to the
anonymous referee for helpful comments. 

The \textit{Fermi} LAT Collaboration acknowledges generous ongoing
support from a number of agencies and institutes that have supported
both the development and the operation of the LAT as well as
scientific data analysis.  These include the National Aeronautics and
Space Administration and the Department of Energy in the United
States, the Commissariat \`a l'Energie Atomique and the Centre
National de la Recherche Scientifique/Institut National de Physique
Nucl\'eaire et de Physique des Particules in France, the Agenzia
Spaziale Italiana and the Istituto Nazionale di Fisica Nucleare in
Italy, the Ministry of Education, Culture, Sports, Science and
Technology (MEXT), High Energy Accelerator Research Organization (KEK)
and Japan Aerospace Exploration Agency (JAXA) in Japan, and the
K.~A.~Wallenberg Foundation, the Swedish Research Council and the
Swedish National Space Board in Sweden.

Additional support for science analysis during the operations phase is
gratefully acknowledged from the Istituto Nazionale di Astrofisica in
Italy and the Centre National d'\'Etudes Spatiales in France.
This work was performed in part under DOE
Contract DE- AC02-76SF00515.
\end{ack}

\appendix

\section{XRT data}

Table~\ref{tab:XRTdata} lists the epochs of the XRT data used in this paper.

\begin{table*}
\caption{XRT data of 41 epochs}
\begin{center}

\scalebox{1}[0.79]{ 
\begin{tabular}{c c c c c} \hline
obs ID &Date (MJD)& Flux ($10^{-9} \rm{erg/cm^2/s}$) & Reduced ${\chi^2_{\rm SED}}^\dag$ \\ \hline
00030352158&55156&$1.2367^{+0.0110}_{-0.0109}$ & $2.01$ \\
00030352161&55176&$0.7883^{+0.0133}_{-0.0130}$ & $2.19$  \\
00030352164&55182&$2.4442^{+0.0239}_{-0.0237}$ & $2.07$  \\
00030352165&55186&$0.8481^{+0.0132}_{-0.0130}$ & $1.25$  \\
00030352167&55190&$1.2015^{+0.0148}_{-0.0146}$ & $1.21$  \\
00030352173&55202&$1.5379^{+0.0178}_{-0.0177}$ & $1.98$   \\
00030352175&55205&$1.1948^{+0.0105}_{-0.0104}$ & $3.41$  \\
00030352176&55207&$0.9499^{+0.0010}_{-0.0099}$ & $1.24$  \\
00030352177&55209&$1.0231^{+0.0132}_{-0.0131}$ & $1.88$  \\
00030352178&55210&$1.8317^{+0.0168}_{-0.0166}$ & $1.58$  \\
00030352179&55211&$2.1356^{+0.0208}_{-0.0208}$ & $3.15$  \\
00030352182&55212&$0.4834^{+0.0106}_{-0.0109}$ & $1.61$  \\
00030352183&55213&$0.6088^{+0.0074}_{-0.0074}$ & $1.87$  \\
00030352185&55215&$1.6785^{+0.0203}_{-0.0202}$ & $1.36$  \\
00030352188&55220&$0.5938^{+0.0124}_{-0.0123}$ & $1.32$  \\
00030352195&55230&$0.4209^{+0.0095}_{-0.0924}$ & $1.25$  \\
00030352198&55235&$0.7029^{+0.0119}_{-0.0118}$ & $1.43$  \\
00030352201&55238&$0.8301^{+0.0123}_{-0.0122}$ & $2.01$  \\
00030352203&55239&$1.7743^{+0.0176}_{-0.0176}$ & $1.58$  \\
00030352213&55252&$1.7707^{+0.0199}_{-0.0198}$ & $0.94$  \\
00030352217&55257&$2.2342^{+0.0213}_{-0.0212}$ & $1.73$  \\
00030352221&55261&$1.0977^{+0.0134}_{-0.0134}$ & $1.49$  \\
00030352230&55272&$0.4402^{+0.0329}_{-0.0300}$ & $0.98$  \\
00030352233&55275&$0.7999^{+0.0108}_{-0.0107}$ & $1.54$  \\
00030352242&55287&$0.7622^{+0.0134}_{-0.0133}$ & $1.24$  \\
00031202007&55290&$1.1001^{+0.0187}_{-0.0186}$ & $1.38$  \\
00031202015&55309&$0.8228^{+0.0149}_{-0.0140}$ & $1.12$  \\
00031202016&55312&$1.1803^{+0.0165}_{-0.0164}$ & $1.24$  \\
00031202019&55318&$1.1946^{+0.0137}_{-0.0137}$ & $1.42$  \\
00031202020&55319&$0.9743^{+0.0127}_{-0.0126}$ & $1.45$  \\
00031202032&55339&$1.3999^{+0.0155}_{-0.0154}$ & $1.44$  \\
00031202060&55568&$0.2592^{+0.0120}_{-0.0110}$ & $1.57$  \\
00031202061&55570&$0.2403^{+0.0078}_{-0.0074}$ & $1.25$  \\
00031202076&55598&$0.1414^{+0.0080}_{-0.0071}$ & $0.84$  \\
00031202087&55623&$0.1741^{+0.0063}_{-0.0060}$ & $3.74$  \\
00031202094&55633&$0.2189^{+0.0083}_{-0.0078}$ & $3.33$  \\
00035014025&56300&$0.1525^{+0.0086}_{-0.0076}$ & $0.99$  \\
00035014032&56304&$0.3149^{+0.0099}_{-0.0094}$ & $1.34$  \\
00035014072&56414&$0.4501^{+0.0055}_{-0.0054}$ & $1.19$  \\
00035014076&56428&$0.2499^{+0.0082}_{-0.0078}$ & $1.14$  \\
00035014108&56708&$0.4167^{+0.0159}_{-0.0102}$ & $1.35$  \\ \hline
\multicolumn{4}{l}{$^\dag$Reduced $\chi^2$ of the SED analysis shown
  in section 5.2}
\end{tabular}
}
\end{center}\label{tab:XRTdata}
\end{table*}

\section{Trace plots}

\begin{figure}
  \begin{center}
    \includegraphics[clip,width=8cm]{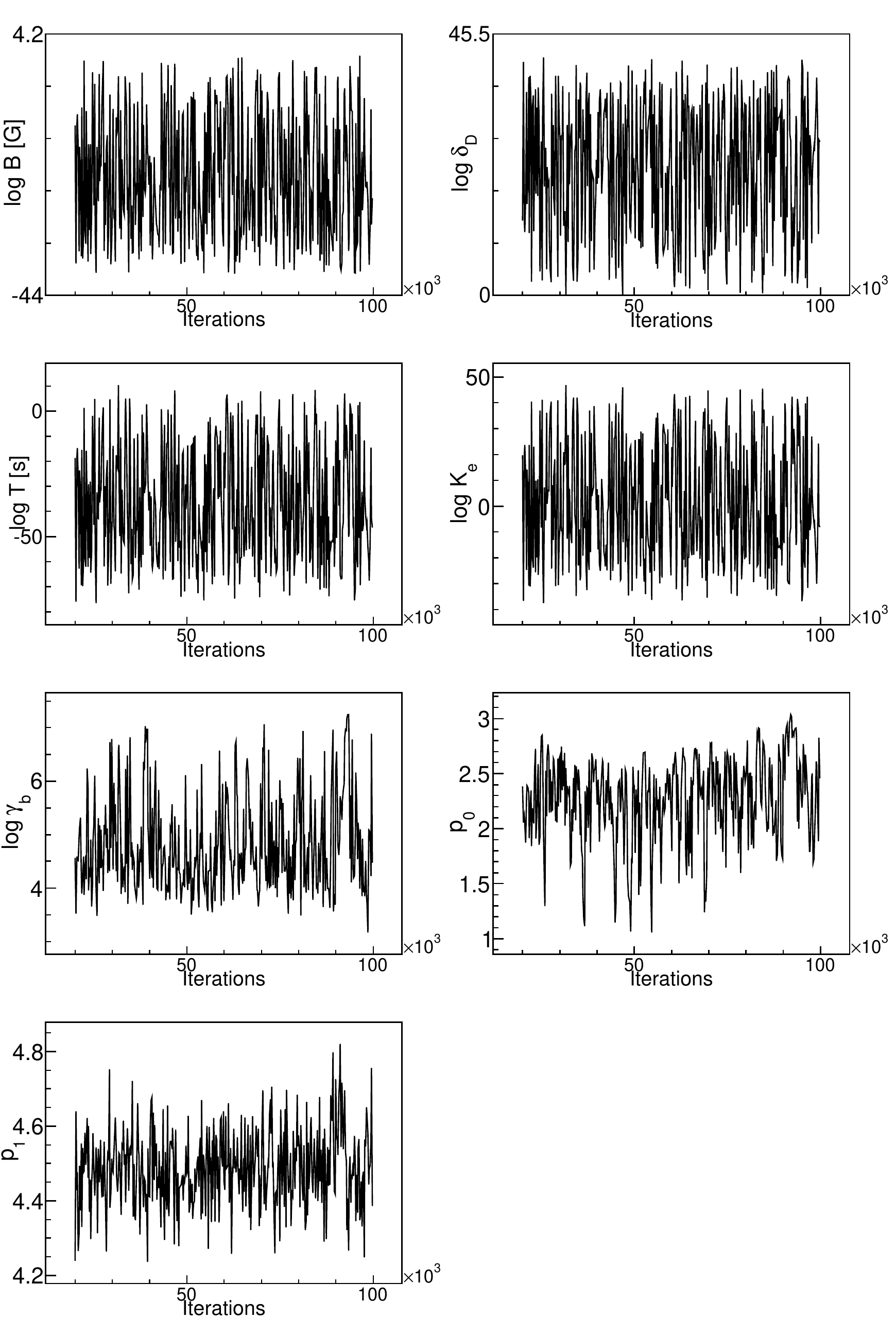}
\end{center}
    \caption{Trace plots of seven parameters, $\log B$, $\log
      \delta_{\rm{D}}$, $\log T$, $\log K_{\rm{e}}$, $\log
      \gamma_{\rm{b}}$, $p_0$, $p_1$, obtained with MCMC and a model
      with non-informative priors. The data on MJD 55598 was used. The
      horizontal and vertical axes indicate the iteration of the MCMC
      algorithm and each parameter value. The algorithm was iterated
      $10^5$ times. We discarded the first $2.0\times 10^4$ steps as
      burn-in.}\label{fig:notjizen_trace}
\end{figure}

\begin{figure}
\begin{center}
\subfigure{\includegraphics[width=7cm]{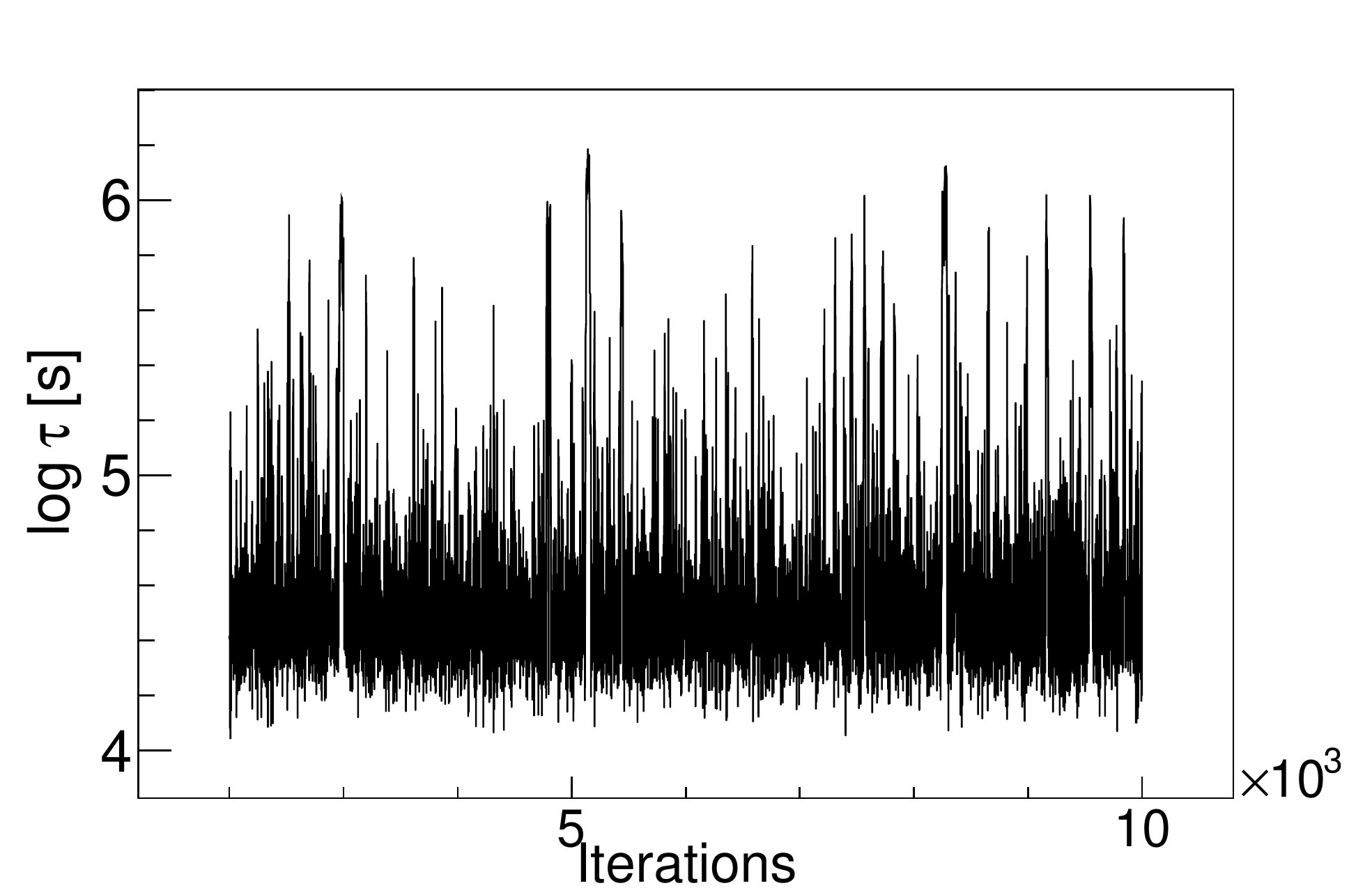}
 \label{fig:fig01left}}
\subfigure{\includegraphics[width=7cm]{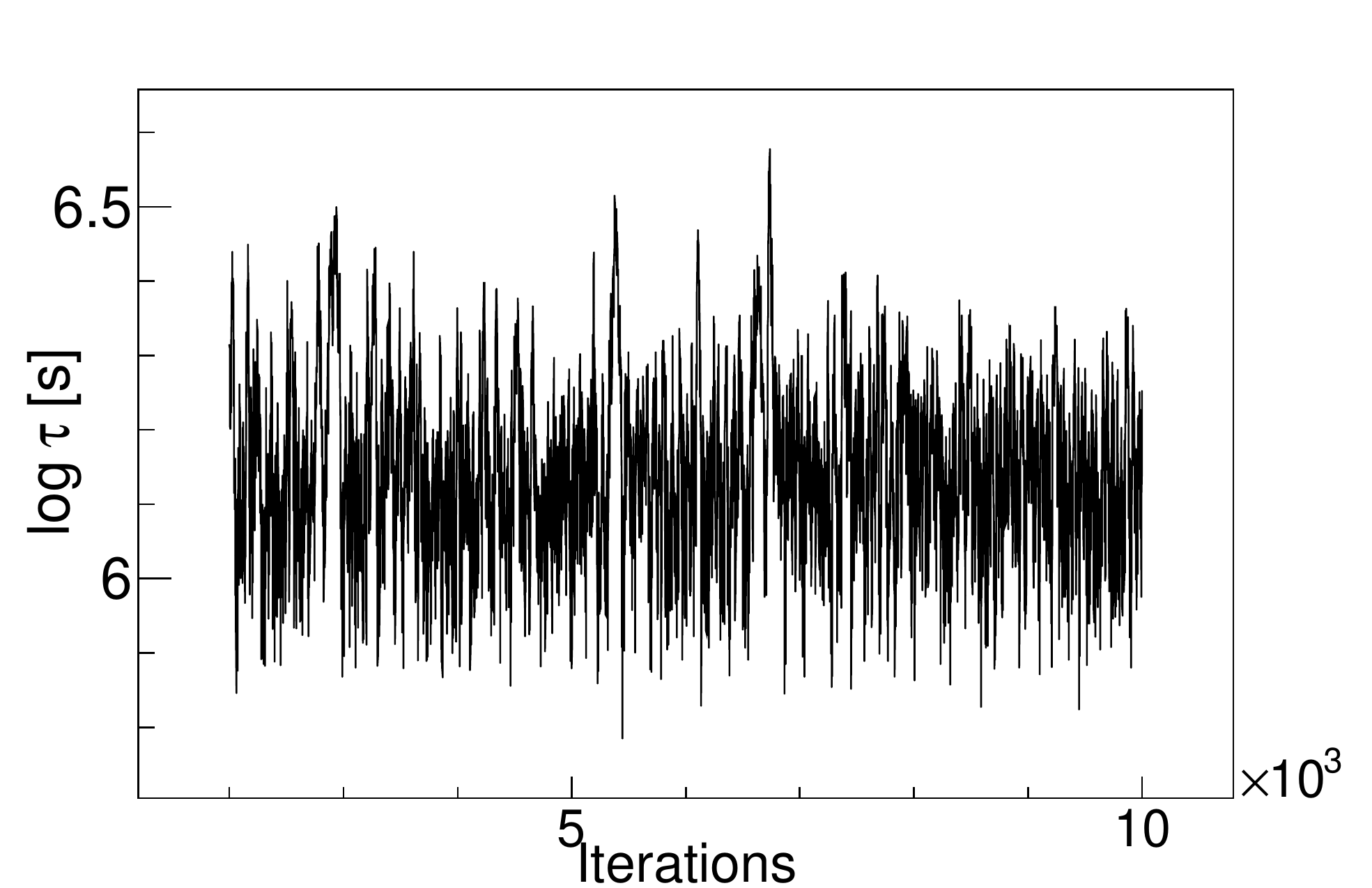}
  \label{fig:fig01right}}
\end{center}
\caption{Trace plots of the time scale $\tau$ obtained with MCMC. The
  left and right panels show the results for the ASCA and
  $\it{Swift}$-XRT data, respectively.}\label{fig:Ttrace}
\end{figure}

Figures~\ref{fig:notjizen_trace} and \ref{fig:Ttrace} show the trace
plots of the MCMC samples for the SED analysis of the data on
MJD~55598 with non-informative prior (section~3.3) and for the OU
process regression of the X-ray light curve (section~4.3). 

We tested the convergence of the MCMC samples by using the Gelman and
Rubin's convergence diagnostic (\cite{Gelman1992};
\cite{Brooks1997}). The method gives the potential scale reduction
factor (PSRF) calculated from the means and variances of multiple
Markov chains with different initial values. We can be confident that
convergence has been achieved if PSRF $< 1.1$ for all parameters. For
example, using the MCMC samples shown in
Figure~\ref{fig:notjizen_trace}, we calculated PSRF to be 1.015,
1.016, 1.017, 1.016, 1.018, 1.049, and 1.018 for the parameter, $\log
B$, $\log \delta_{\rm D}$, $\log T$, $\log K_{\rm e}$, $\log
\gamma_{\rm b}$, $p_0$, and $p_1$, respectively, from five chains each
of which has $8\times 10^4$ samples. We confirmed that the PSRF
criterion was satisfied in all cases in this paper. 

\clearpage

\begin{thebibliography}{}

\bibitem[{Abdo} et~al.(2011)]{Abdo2011}
  {Abdo}, A.~A., {Ackermann}, M., {Ajello}, M., {Baldini}, L., {Ballet}, J.,
  {Barbiellini}, G., {Bastieri}, D., {Bechtol}, K., {et~al.}\ 2011, \apj, 736,
  131

\bibitem[Acero et~al.(2015)]{Acero2015}
  Acero, F., Ackermann, M., Ajello, M., Albert, A., Atwood, W., Axelsson, M.,
  Baldini, L., Ballet, J., {et~al.}\ 2015, The Astrophysical Journal Supplement
  Series, 218, 23

\bibitem[{Acero} et~al.(2016)]{Acero2016}
  {Acero}, F., {Ackermann}, M., {Ajello}, M., {Albert}, A., {Baldini}, L.,
  {Ballet}, J., {Barbiellini}, G., {Bastieri}, D., {et~al.}\ 2016, \apjs, 223,
  26

\bibitem[{Aielli} et~al.(2010)]{Aielli2010}
  {Aielli}, G., {Bacci}, C., {Bartoli}, B., {Bernardini}, P., {Bi}, X.~J.,
  {Bleve}, C., {Branchini}, P., {Budano}, A., {et~al.}\ 2010, \apjl, 714, L208

\bibitem[{Ar{\'e}valo} et~al.(2006)]{Arevalo2006}
  {Ar{\'e}valo}, P., {Papadakis}, I.~E., {Uttley}, P., {McHardy}, I.~M., \&
  {Brinkmann}, W.\ 2006, \mnras, 372, 401


\bibitem[{Atwood} et~al.(2009)]{Atwood2009}
  {Atwood}, W.~B., {Abdo}, A.~A., {Ackermann}, M., {Althouse}, W., {Anderson},
  B., {Axelsson}, M., {Baldini}, L., {Ballet}, J., {et~al.}\ 2009, \apj, 697,
  1071

\bibitem[{Atwood} et~al.(2013)]{Atwood2013}
  {Atwood}, W.~B., {Baldini}, L., {Bregeon}, J., {Bruel}, P., {Chekhtman}, A.,
  {Cohen-Tanugi}, J., {Drlica-Wagner}, A., {Granot}, J., {et~al.}\ 2013, \apj,
  774, 76

\bibitem[Bartoli et~al.(2011)]{Bartoli2011a}
  Bartoli, B., Bernardini, P., Bi, X., Bleve, C., Bolognino, I., Branchini, P.,
  Budano, A., Melcarne, A.~C., {et~al.}\ 2011, The Astrophysical Journal, 734,
  110

\bibitem[{Bartoli} et~al.(2016)]{Bartoli2016}
  {Bartoli}, B., {Bernardini}, P., {Bi}, X.~J., {Cao}, Z., {Catalanotti}, S.,
  {Chen}, S.~Z., {Chen}, T.~L., {Cui}, S.~W., {et~al.}\ 2016, \apjs, 222, 6

\bibitem[B{\l}a{\.z}ejowski et~al.(2005)]{Blazejowski2005}
  B{\l}a{\.z}ejowski, M., Blaylock, G., Bond, I., Bradbury, S., Buckley, J.,
  Carter-Lewis, D., Celik, O., Cogan, P., {et~al.}\ 2005, The Astrophysical
  Journal, 630, 130

\bibitem[{B{\"o}ttcher}(2007)]{Bottcher2007}
  {B{\"o}ttcher}, M.\ 2007, \apss, 309, 95

\bibitem[B{\"o}ttcher et~al.(2013)]{Bottcher2013}
  B{\"o}ttcher, M., Reimer, A., Sweeney, K., \& Prakash, A.\ 2013, The
  Astrophysical Journal, 768, 54

\bibitem[{Brooks} \& {Gelman}(1997)]{Brooks1997}
  {Brooks}, S.~P., {Gelman}, A.\ 1997. Journal of Computational and
  Graphical Statistics, 7, 434
  
\bibitem[Burke et~al.(1994)]{Burke1994}
  Burke, B., Mountain, R., Daniels, P., Cooper, M., \& Dolat, V.\ 1994, IEEE
  transactions on nuclear science, 41, 375

\bibitem[{Burrows} et~al.(2005)]{Burrows2005}
  {Burrows}, D.~N., {Hill}, J.~E., {Nousek}, J.~A., {Kennea}, J.~A., {Wells},
  A., {Osborne}, J.~P., {Abbey}, A.~F., {Beardmore}, A., {et~al.}\ 2005, \ssr,
  120, 165

\bibitem[Cardelli et~al.(1989)]{Cardelli1989}
  Cardelli, J.~A., Clayton, G.~C., \& Mathis, J.~S.\ 1989, The Astrophysical
  Journal, 345, 245

\bibitem[{Ding} et~al.(2017)]{Ding2017}
  {Ding}, N., {Zhang}, X., {Xiong}, D.~R., \& {Zhang}, H.~J.\ 2017, \mnras,
  464, 599

\bibitem[{Donnarumma} et~al.(2009)]{Donnarumma2009}
  {Donnarumma}, I., {Vittorini}, V., {Vercellone}, S., {del Monte}, E.,
  {Feroci}, M., {D'Ammando}, F., {Pacciani}, L., {Chen}, A.~W., {et~al.}\ 2009,
  \apjl, 691, L13

\bibitem[Finke et~al.(2008)]{Finke2008}
  Finke, J.~D., Dermer, C.~D., \& B{\"o}ttcher, M.\ 2008, The Astrophysical
  Journal, 686, 181

\bibitem[Fossati et~al.(2008)]{Fossati2008}
  Fossati, G., Buckley, J., Bond, I., Bradbury, S., Carter-Lewis, D., Chow, Y.,
  Cui, W., Falcone, A., {et~al.}\ 2008, The Astrophysical Journal, 677, 906

\bibitem[{Fossati} et~al.(1998)]{Fossati}
  {Fossati}, G., {Maraschi}, L., {Celotti}, A., {Comastri}, A., \&
  {Ghisellini}, G.\ 1998, \mnras, 299, 433

\bibitem[{Gehrels} et~al.(2004)]{Gehrels2004}
  {Gehrels}, N., {Chincarini}, G., {Giommi}, P., {Mason}, K.~O., {Nousek},
  J.~A., {Wells}, A.~A., {White}, N.~E., {Barthelmy}, S.~D., {et~al.}\ 2004,
  \apj, 611, 1005

\bibitem[{Gelman} \& {Rubin}(1992)]{Gelman1992}
  {Gelman}, A., {Rubin}, D~B.\ 1992, Statistical Science, 7, 457
  
\bibitem[{Ghisellini} et~al.(2009)]{Ghisellini2009}
  {Ghisellini}, G., {Tavecchio}, F., \& {Ghirlanda}, G.\ 2009, \mnras, 399,
  2041
  
\bibitem[{Haario} et~al.(2001)]{Haario2001}
  {Haario}, H., {Saksman}, E., \& {Tamminen}, J.\ 2001, Bernoulli, 7,
  223
  
\bibitem[{Itoh} et~al.(2015)]{itoh2015}
  {Itoh}, R., {Fukazawa}, Y., {Tanaka}, Y.~T., {Kawabata}, K.~S., {Takaki}, K.,
  {Hayashi}, K., {Uemura}, M., {Ui}, T., {et~al.}\ 2015, \pasj, 67, 45

\bibitem[{Kawabata} et~al.(2008)]{Kawabata2008}
  {Kawabata}, K.~S., {Nagae}, O., {Chiyonobu}, S., {Tanaka}, H., {Nakaya}, H.,
  {Suzuki}, M., {Kamata}, Y., {Miyazaki}, S., {et~al.}\ 2008, in Ground-based
  and Airborne Instrumentation for Astronomy II Vol.~7014 of \procspie \ 70144L

\bibitem[Kelly et~al.(2009)]{Kelly2009}
  Kelly, B.~C., Bechtold, J., \& Siemiginowska, A.\ 2009, The Astrophysical
  Journal, 698, 895

\bibitem[Kelly et~al.(2011)]{Kelly2011}
  Kelly, B.~C., Sobolewska, M., \& Siemiginowska, A.\ 2011, The Astrophysical
  Journal, 730, 52

\bibitem[Kino et~al.(2002)]{Kino2002}
  Kino, M., Takahara, F., \& Kusunose, M.\ 2002, The Astrophysical Journal,
  564, 97

\bibitem[{Lockman}, {Savage}(1995)]{Lockman1995}
  {Lockman}, F.~J. \& {Savage}, B.~D.\ 1995, \apjs, 97, 1

\bibitem[Makishima et~al.(1996)]{Makishima1996}
  Makishima, K., Tashiro, M., Ebisawa, K., Ezawa, H., Fukazawa, Y., Gunji, S.,
  Hirayama, M., Idesawa, E., {et~al.}\ 1996, Publications of the Astronomical
  Society of Japan, 48, 171

\bibitem[{McHardy} et~al.(2007)]{McHardy2007}
  {McHardy}, I.~M., {Ar{\'e}valo}, P., {Uttley}, P., {Papadakis}, I.~E.,
  {Summons}, D.~P., {Brinkmann}, W., \& {Page}, M.~J.\ 2007, \mnras, 382, 985

\bibitem[Ohashi et~al.(1996)]{Ohashi1996}
  Ohashi, T., Ebisawa, K., Fukazawa, Y., Hiyoshi, K., Horii, M., Ikebe, Y.,
  Ikeda, H., Inoue, H., {et~al.}\ 1996, Publications of the Astronomical
  Society of Japan, 48, 157

\bibitem[{Paggi} et~al.(2009)]{Paggi2009}
  {Paggi}, A., {Cavaliere}, A., {Vittorini}, V., \& {Tavani}, M.\ 2009, \aap,
  508, L31

\bibitem[{Poole} et~al.(2008)]{Poole2008}
  {Poole}, T.~S., {Breeveld}, A.~A., {Page}, M.~J., {Landsman}, W., {Holland},
  S.~T., {Roming}, P., {Kuin}, N.~P.~M., {Brown}, P.~J., {et~al.}\ 2008,
  \mnras, 383, 627

\bibitem[{Punch} et~al.(1992)]{Punch1992}
  {Punch}, M., {Akerlof}, C.~W., {Cawley}, M.~F., {Chantell}, M., {Fegan},
  D.~J., {Fennell}, S., {Gaidos}, J.~A., {Hagan}, J., {et~al.}\ 1992, \nat,
  358, 477

\bibitem[Robbins \& Monro(1951)]{Robbins1951}
  Robbins, H. \& Monro, S.\ 1951, Statistics, 22, 400

\bibitem[{Roberts} et~al.(1997)]{Roberts1997}
  {Roberts}, G.~O., {Gelman}, A., {Gilks}, W.~R.\ 1997, Annals of
  Applied Probability, 7, 110
  
\bibitem[{Roming} et~al.(2005)]{Roming2005}
  {Roming}, P.~W.~A., {Kennedy}, T.~E., {Mason}, K.~O., {Nousek}, J.~A., {Ahr},
  L., {Bingham}, R.~E., {Broos}, P.~S., {Carter}, M.~J., {et~al.}\ 2005, \ssr,
  120, 95

\bibitem[{Schlafly} \& {Finkbeiner}(2011)]{Schlafly2011}
  {Schlafly}, E.~F. \& {Finkbeiner}, D.~P.\ 2011, \apj, 737, 103

\bibitem[{Shukla} et~al.(2012)]{Shukla2012}
  {Shukla}, A., {Chitnis}, V.~R., {Vishwanath}, P.~R., {Acharya}, B.~S.,
  {Anupama}, G.~C., {Bhattacharjee}, P., {Britto}, R.~J., {Prabhu}, T.~P.,
  {Saha}, L., \& {Singh}, B.~B.\ 2012, \aap, 541, A140


\bibitem[{Sobolewska} et~al.(2014)]{Sobolewska2014}
  {Sobolewska}, M.~A., {Siemiginowska}, A., {Kelly}, B.~C., \& {Nalewajko}, K.\
  2014, \apj, 786, 143

\bibitem[{Spada} et~al.(2001)]{Spada2001}
  {Spada}, M., {Ghisellini}, G., {Lazzati}, D., \& {Celotti}, A.\ 2001, \mnras,
  325, 1559

\bibitem[Tanaka et~al.(1994)]{Tanaka1994}
  Tanaka, Y., Inoue, H., \& Holt, S.~S.\ 1994, Publications of the Astronomical
  Society of Japan, 46, L37

\bibitem[Tavecchio et~al.(1998)]{Tavecchio1998}
  Tavecchio, F., Maraschi, L., \& Ghisellini, G.\ 1998, The Astrophysical
  Journal, 509, 608

\bibitem[{Tramacere} et~al.(2009)]{Tramacere2009}
  {Tramacere}, A., {Giommi}, P., {Perri}, M., {Verrecchia}, F., \& {Tosti}, G.\
  2009, \aap, 501, 879

\bibitem[{Tucker}(1975)]{Tucker1975}
  {Tucker}, W.\ 1975, {Radiation processes in astrophysics} (Cambridge, Mass., MIT Press)

\bibitem[Uhlenbeck, Ornstein(1930)]{Uhlenbeck1930}
  Uhlenbeck, G.~E. \& Ornstein, L.~S.\ 1930, Physical review, 36, 823

\bibitem[{Urry}, {Padovani}(1995)]{Urry}
  {Urry}, C.~M. \& {Padovani}, P.\ 1995, \pasp, 107, 803

\bibitem[{Vianello} et~al.(2015)]{gia15MMML}
  {Vianello}, G., {Lauer}, R. J., {Younk}, P., {Tibaldo}, L.,
  {Burgess}, J. M., {Ayala}, H., {Harding}, P., {Hui}, M., {Omodei},
  N., \& {Zhou}, H.\ 2015, Proceedings of the 34th International
  Cosmic Ray Conference (ICRC2015), 1042 (arXiv:1507.08343)

\bibitem[Yamashita et~al.(1997)]{Yamashita1997}
  Yamashita, A., Dotani, T., Bautz, M., Crew, G., Ezuka, H., Gendreau, K.,
  Kotani, T., Mitsuda, K., {et~al.}\ 1997, IEEE Transactions on Nuclear
  Science, 44, 847

\bibitem[Yan et~al.(2013)]{Dahai2013}
  Yan, D., Zhang, L., Yuan, Q., Fan, Z., \& Zeng, H.\ 2013, The Astrophysical
  Journal, 765, 122

\end{thebibliography}

\end{document}